\documentclass[sigconf]{acmart}
\AtBeginDocument{%
	}

\acmYear{2023}
\setcopyright{none}
\acmConference[ACSAC '23]{Annual Computer Security Applications 
Conference}{December 4--8, 2023}{Austin, TX, USA}
\acmBooktitle{Annual Computer Security Applications Conference (ACSAC 
'23), December 4--8, 2023, Austin, TX, USA}
\acmDOI{}
\acmPrice{}
\acmISBN{}

\usepackage{amsmath,amsfonts}
\usepackage{algorithmic}
\usepackage{graphicx}
\usepackage{textcomp}
\usepackage{xcolor}
\usepackage{xspace}
\usepackage{microtype}
\usepackage[]{natbib}

\usepackage{multirow}
\usepackage{balance}
\usepackage{booktabs}
\usepackage{tabularx}
\usepackage[]{threeparttable}
\usepackage[font=small]{subcaption}
\usepackage[font={small}]{caption}

\BeforeBeginEnvironment{tabularx}{\footnotesize}
\BeforeBeginEnvironment{tabular}{\footnotesize}

\newcolumntype{R}{>{\raggedleft\arraybackslash}X}

\usepackage{enumitem}

\definecolor{cbone}  {HTML}{006BA4} %
\definecolor{cbtwo}  {HTML}{FF800E} %
\definecolor{cbfour} {HTML}{595959} %
\definecolor{cbfive} {HTML}{5F9ED1} %
\definecolor{cbnine} {HTML}{FFBC79} %
\definecolor{cbten}  {HTML}{CFCFCF} %

\definecolor{tubsredSec}{cmyk}{0.0,1.00,0.6,0.6}
\definecolor{tubsredPrim}{cmyk}{0.1,1.0,0.8,0.0}

\colorlet{primarycolor}{cbone}
\colorlet{primaryshaded}{cbfive}
\colorlet{secondarycolor}{cbtwo}
\colorlet{secondaryshaded}{cbnine}
\colorlet{tertiarycolor}{cbfour}
\colorlet{tertiaryshaded}{cbten}

\definecolor{sunygreen}{RGB}{0,166,77}
\definecolor{darkgreentone}{RGB}{60,179,113}

\renewcommand{\paragraph}[1]{{\vskip 6pt \noindent\textbf{#1.} }}
\newcommand{\minipara}[1]{\emph{#1.~}}

\newcommand{\srcimg}{\ensuremath{S}\xspace}
\newcommand{\tarimg}{\ensuremath{T}\xspace}
\newcommand{\attimg}{\ensuremath{A}\xspace}
\newcommand{\outimg}{\ensuremath{D}\xspace}
\newcommand{\deltaS}{\ensuremath{\Delta}\xspace}
\newcommand{\scalefunc}{\ensuremath{\mathrm{scale}}}
\newcommand{\downscalefunc}{\ensuremath{\mathrm{downscale}}}
\newcommand{\upscalefunc}{\ensuremath{\mathrm{upscale}}}
\newcommand{\filter}{\ensuremath{\mathcal{V}\xspace}}

\newcommand{\sr}{\ensuremath{\beta}\xspace} %
\newcommand{\goalA}{O1\xspace} %
\newcommand{\goalB}{O2\xspace} %

\newcommand{\ie}{i.e.,\xspace} %

\newcommand{\adaptiveattimg}{\ensuremath{\tilde{A}}\xspace} 
\newcommand{\suppressfactor}{\ensuremath{f_s}\xspace} 
\newcommand{\addpeakfactor}{\ensuremath{f_a}\xspace}

\newcommand{\Fourier}{\ensuremath{F}\xspace} 
 
\newcommand{\Tm}{\ensuremath{\sr_m}\xspace} 
\newcommand{\Tn}{\ensuremath{\sr_n}\xspace} 

\newcommand{\w}{\ensuremath{w}\xspace} %
\newcommand{\window}{\ensuremath{\omega}\xspace} %
\newcommand{\stride}{\ensuremath{s}\xspace} %

\usepackage{cleveref} %

\newcommand{\thisw}{\textasteriskcentered~~} %
\newcommand{\thiswn}{\phantom{\textasteriskcentered~~}} %
\newcommand{\thisws}{\textasteriskcentered~} %

\newcommand{\appref}[1]{\hyperref[#1]{Appendix~\ref{#1}}}

\usepackage{catchfile}

\usepackage{fancyhdr}

\begin{document}
\title{On the Detection of Image-Scaling Attacks in Machine Learning}

\author{Erwin Quiring}
\affiliation{
	\institution{ICSI}
	\city{Berkeley}
	\country{United States}
}
\affiliation{%
	\institution{Ruhr University Bochum}
	\city{Bochum}
	\country{Germany}
	\vspace{1.5cm}
}

\author{Andreas Müller}
\affiliation{%
	\institution{Ruhr University Bochum}
	\city{Bochum}
	\country{Germany}
}

\author{Konrad~Rieck}
\affiliation{%
	\institution{TU Berlin}
	\city{Berlin}
	\country{Germany}
}

\renewcommand{\shortauthors}{Quiring et al.}

\begin{abstract}
  Image scaling is an integral part of machine learning and computer
  vision systems. Unfortunately, this preprocessing step is vulnerable to
  so-called image-scaling attacks where an attacker makes 
  unnoticeable changes to an image 
  so that it becomes a new image after scaling.
  This opens up new ways for attackers to control the 
  prediction or to improve poisoning and backdoor attacks.
  While effective
  techniques exist to prevent scaling attacks, their detection has not
  been rigorously studied yet. Consequently, it is currently not
  possible to reliably spot these attacks in practice.

  This paper presents the first
  in-depth systematization and analysis of detection methods for image-scaling
  attacks. We identify two general detection paradigms and
  derive novel methods from them that are simple in
  design yet significantly outperform previous work.
  We demonstrate the efficacy of these methods in a comprehensive
  evaluation with all major learning platforms and scaling
  algorithms. First, we show that image-scaling attacks modifying the
  entire scaled image can be reliably detected even under an adaptive
  adversary. Second, we find that our methods provide strong detection
  performance even if only minor parts of the image are manipulated.
  As a result, we can introduce a novel protection layer against
  image-scaling attacks. %
\end{abstract}


\maketitle

\fancypagestyle{firststyle}
{
	\fancyhf{}
	\chead{\small\textit{-------------------------------------- 
Accepted at the Annual Computer Security Applications Conference (ACSAC) 2023
			--------------------------------------}}	
	\renewcommand{\headrulewidth}{0pt} }
\thispagestyle{firststyle}

\section{Introduction}

\begin{figure}[b]
	\centering      
	\includegraphics{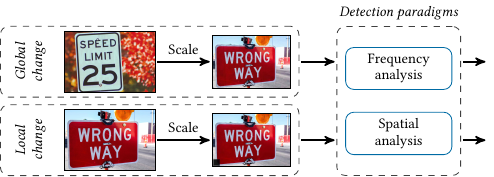}
	\vspace{0.0cm}
	\caption{Image-scaling attacks: global modification (top) and 
	local modification (bottom). Note the small box (backdoor) on the 
		lower left that appears. Both attacks can be \emph{detected} 
		by using a frequency or spatial analysis.}
	\label{fig:intro_example}
\end{figure}

Image scaling is a ubiquitous preprocessing step in many machine
learning and computer vision systems. Before an image is fed to a
learning model for inference, it is usually scaled down to fixed
dimensions. For example, the popular neural networks
VGG19~\citep{SimZis14} and \mbox{ResNet}~\citep{HeZhaRen+16} for
object recognition expect fixed inputs of $224 \times 224$ pixels.
While an extensive body of research has explored vulnerabilities in
learning models~\citep{PapMcSin+18, BigRol18}, the attack surface of
preprocessing has received little attention so far.  An exception is
recent work on \mbox{\emph{image-scaling 
attacks}}~\citep{XiaCheShe+19, QuiKleArp20}, a novel class of attacks
that enable an adversary to tamper with the result of the scaling
process (see~\autoref{fig:intro_example}). These 
attacks exploit
that most scaling algorithms process only a minor fraction of the
pixels in an image, so that a few perturbations allow for full control
of its scaled version~\citep{QuiKleArp20}.

In contrast to other security threats to machine learning, image-scaling
attacks are agnostic to the employed learning models.  Successful
attacks only require knowledge about the scaling algorithm and the
target dimensions.
Compared to the parameters of a neural network, these details are
limited in the number of possible configurations and 
can also be inferred through remote queries to the model~\citep{XiaCheShe+19}.
As a result, image-scaling attacks pose a notable threat to practical
systems: They enable misleading classifiers without access to the 
learning model and allow hiding backdoor triggers or poisoning 
attacks in training
data~\citep{QuiRie20}.  \autoref{fig:intro_example} illustrates both
cases.  In the top row, the adversary misleads the classification by
changing the entire image during scaling. In the bottom row, the
adversary induces local changes in the lower left corner of the scaled
image (black square). If this modification is performed on training
data, it allows concealing an otherwise noticeable backdoor trigger.
Hence, there is a need for effective
safeguards that complement existing security mechanisms for machine
learning. %

Two defense strategies have been explored for addressing this threat:
\emph{prevention} and \emph{detection}. In the first case, robust
scaling algorithms or specific defense filters are applied to input
images to prevent attacks. While these defenses are provably
effective, as demonstrated by \citet{QuiKleArp20}, they can only
prevent but not detect attacks. 
However, detection can be necessary in various cases:
First, we can spot on-going attacks. This allows scanning image 
collections for attacks or identifying the adversary.
Second, robust algorithms like area scaling are slower compared to 
vulnerable algorithms, so that a one-time check might be preferred in 
real-time settings. Third, detection allows protecting proprietary
learning systems where components cannot be changed. 

The detection of image-scaling attacks, however, has not been
rigorously studied~yet. 
In this paper, we address this research gap and present the first
in-depth systematization 
and analysis of detection methods for image-scaling
attacks. We identify two general paradigms that underlie existing
detection approaches: \mbox{\emph{frequency analysis}} and 
\emph{spatial 
analysis}.  In the first paradigm, the detection is based on 
searching for conspicuous traces in the frequency spectrum. 
In the second paradigm, the pixels are analyzed, either by 
leveraging the \emph{adversarial} modification or the 
remaining \emph{clean} pixels for further analysis.

We derive novel detection methods for these paradigms that are simple
yet effective in spotting image-scaling attacks.  Our
methods significantly outperform previous approaches, as we can
reduce heuristic elements and precisely pinpoint the detectable
characteristics of image-scaling attacks.

We demonstrate the efficacy of our methods in a comprehensive
evaluation.  In contrast to prior work, we study the detection
performance with a diverse evaluation setup, including all major
learning platforms, scaling algorithms, a large dataset, static \&
adaptive adversaries, and different attack scenarios such as global 
\& local changes.
If the entire scaled image is modified, both paradigms allow a
detection rate between 99\% and 100\%. Our frequency approaches are
particularly effective with a perfect detection rate.  In the
challenging scenario where only a local area is manipulated, previous
methods fail while the newly derived approaches remain effective.  The
frequency paradigm enables a detection rate of~90\%.  The spatial
paradigm allows detecting at least three out of four attacks. Under an
adaptive attacker, however, the paradigms' effectiveness flips: The
frequency paradigm---the strongest in the static attack---does not
withstand an adaptive attack while the spatial analysis remains
robust.  As a result, we conclude that both paradigms should be used
in combination to complement each other.

\paragraph{Contributions} In summary, our contributions are as 
follows:

\begin{enumerate}
\setlength{\itemsep}{3pt}

\item {\em Systematization of detection.} We present the
  first systematic analysis of detection methods for image-scaling
  attacks where we identify two general paradigms. %

\item {\em Novel approaches to detection. } Based on our analysis, we
  derive novel detection approaches that %
  significantly outperform all approaches from previous work.

\item {\em Comprehensive evaluation. } We empirically investigate the
  performance of all detection approaches in a comprehensive
  evaluation that covers all major learning platforms and scaling
  algorithms.

\item {\em Different attack models. } Moreover, we carefully examine
  the limits of all detection approaches by experimenting with
  different attack scenarios, such as global \& local changes, 
  and with both static \& adaptive attackers.

\end{enumerate}
To foster further research in this area, our code is publicly 
available at 
\textcolor{primarycolor}{\url{https://github.com/EQuiw/2023-detection-scalingattacks}}.

\paragraph{Roadmap} We introduce image-scaling
attacks in \autoref{sec:background}. The detection paradigms
with our new methods are presented in
\autoref{sec:detection-methods} and the evaluation of
their performance is given in \autoref{sec:evaluation}. Adaptive 
attacks are evaluated separately in \autoref{sec:adaptiveattack}. 
Finally, \autoref{sec:relatedwork} discusses related work and 
\autoref{sec:conclusion} concludes the paper.

\section{Background} \label{sec:background} 
We start by introducing the background on image-scaling attacks before
presenting our novel detection approaches.

\subsection{Preprocessing in Machine Learning}
When solving tasks of computer vision, image data is typically 
normalized and preprocessed before features are extracted and learning
models are applied.
In particular, \emph{image scaling} is a widely used preprocessing
step to bring images to a normalized form. Most learning algorithms
expect a fixed input size and thus images with different or larger
dimensions need to be scaled. For example, the deep neural networks
VGG19~\citep{SimZis14} and \mbox{ResNet}~\citep{HeZhaRen+16} require
inputs of $224 \times 224$ pixels.
Due to this frequent preprocessing, major machine-learning frameworks
directly integrate different scaling algorithms. For instance, Caffe
employs the image processing library \emph{OpenCV}, PyTorch uses the
library \emph{Pillow}, and TensorFlow has its own implementation
called \emph{tf.image}.  Following prior
work~\cite{XiaCheShe+19}, we thus focus our analysis on the libraries
OpenCV, Pillow, and tf.image with their respective implementations of
scaling algorithms.

\subsection{Image-Scaling Attacks}
\label{subsec:background-scaling-attacks}
Interestingly, the downscaling of images leads to a considerable
attack surface in learning-based systems. By carefully modifying
particular pixels, it becomes possible to control the output of the
scaling algorithms and thus to change the content of the scaled image. In
the following, we recap the current state-of-the-art of these
\emph{image-scaling attacks}~\cite{XiaCheShe+19, QuiKleArp20}.

\begin{figure}[t]
	\centering
	\includegraphics{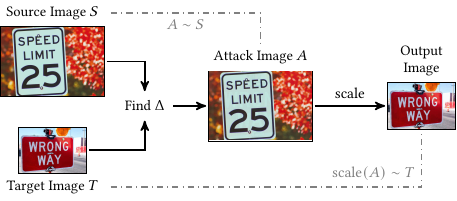}
	\vspace{-0.9em}
	\caption{Principle of image-scaling attacks: 
		The adversary finds a minimal modification \deltaS of \srcimg 
		such that the modified image~$\attimg = (\srcimg + \deltaS)$ 
		still looks like~\srcimg, but downscales to \tarimg.}
	\label{fig:scaling_attack_example}
\end{figure}

\autoref{fig:scaling_attack_example} exemplifies the principle of the 
attack. Given a source image~\srcimg, the adversary 
tries to find a minimal perturbation~\deltaS, so that the downscaling 
of the modified image~$\attimg = (\srcimg + \deltaS)$ produces an 
output image, $\scalefunc(\attimg)$, that matches the adversary's 
target image~\tarimg.
This attack can be modeled as a quadratic optimization 
problem:
\begin{align}
	&\min ( \Vert \deltaS \Vert_2^2 ) \nonumber \\
	\mathrm{s.t.}\quad &\Vert \scalefunc(\srcimg + \deltaS) - \tarimg
	\Vert_{\infty} \leqslant \epsilon  
	\text{ and } \attimg \in \mathit{R} ,
	\label{eq:opti_problem_basic}
\end{align}
where the interval $\mathit{R}$ is the allowed pixel range, as for
example, $\mathit{R}=[0,255]$ for 8-bit images.

To deceive the victim, a successful attack has to fulfill two
objectives: First, the scaled image,
$\scalefunc(\attimg) = \scalefunc(\srcimg + \deltaS)$, needs to match
the target image~\tarimg, \ie $\scalefunc(\attimg) \sim \tarimg$.
Second, the attack image should be indistinguishable from the source
image, \ie $\attimg \sim \srcimg$.  As a result, the adversary gets an
attack image~\attimg that looks identical to the source \srcimg but
changes to the target~\tarimg after downscaling.

This attack is independent of the training data, extracted features,
and employed learning model.  The adversary only needs to know the
used scaling algorithm and the target size of the scaled image.  In
practice, this knowledge might not be difficult to obtain.  Publicly
available deep neural networks are often re-used through transfer
learning.  %
Moreover, the number of possible scaling configurations in the
libraries is limited.  In some settings, the knowledge about the
algorithm is not even required. OpenCV and TensorFlow, for example,
are effectively using nearest scaling when the scaling factor is an
(uneven) integer~\cite{QuiKleArp20}.  This simplifies an attack if the
adversary can choose the size of \srcimg.  Finally, even without any
knowledge, the scaling setup can be deduced through black-box
queries~\citep{XiaCheShe+19}.

\paragraph{Threat Scenario}
By attacking the preprocessing, the input is manipulated at the
very beginning of the learning pipeline. An adversary can thus
efficiently control any subsequent steps, which enables and simplifies
different attack strategies.
First, during training time, the attacker can conceal poisoning and
backdoor attacks: The training-data modifications become 
visible only after downscaling~\cite{QuiRie20}. This
alleviates the shortcoming of attacks
that leave visible artifacts in images~\citep[][]{GuDolGar17,
  YaoLiZhe+19, ShaHuaNaj+18}. 
The combination is useful, for example, if backdoors are 
used in the 
physical world. At training time, scaling attacks hide the trigger 
that can be later activated in the physical world.
Second, during test time, the adversary can control the
predictions of a learning model---without modifying the training data
or model. The adversary uses the scaling attack, so that the 
downscaling leads to an image of the target class.

Note that scaling attacks allow a similar attack as adversarial 
examples by causing a misclassification. However, they do \emph{not} 
depend on the learning model or features, since the scaling stage 
produces a perfect image of the target class. Scaling attacks would 
succeed even if neural networks were robust against adversarial 
examples.

Attacks can be realized with varying degrees of modification. We 
differentiate two scenarios to understand the detection capabilities:
\begin{itemize}

\setlength{\itemsep}{4pt}

\item \textit{Global modification.} The adversary chooses a target
  image with an arbitrary, unrelated content. The input~\srcimg and
  target~\tarimg have no relation to each other.  This is the most
  severe attack, as any target class can be chosen.

\item \textit{Local modification.}  The source and target image are
  identical, except for a limited area. We study backdoors as
  example for this category where only a small trigger is added~(see
  \autoref{fig:appendix_eval_backdoor_examples}). A detection is 
  challenging due to the small changes.
\end{itemize}

When studying the global scenario, we also consider an \emph{overlay 
scenario} where the scaling attack only partially creates the target 
image~\tarimg. Here, we blend \tarimg into the downscaled 
source image to create a novel target image. This scenario allows 
us to study an attacker who does not fully embed the 
target image into the scaling~output.

\begin{figure}[b]
	\centering
	\includegraphics{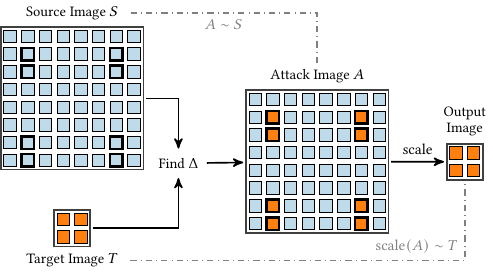}
	\vspace{-1.0em}
	\caption{Simplified illustration of the root cause of 
	image-scaling attacks: The scaling algorithm considers only a few 
	pixels in an input~\srcimg (visualized by a thicker box). 
	The adversary has to change only these
	pixels to control the downscaling output. This added noise is not 
	noticeable.}
	\label{fig:scaling_attack_rootcause}
\end{figure}

\paragraph{Root Cause}
Image-scaling attacks are possible because scaling algorithms do
not process all pixels equally~\citep{QuiKleArp20}. Depending
on the algorithm and scaling ratio, many pixels in \srcimg have
limited or even no impact on the scaled output.  An adversary can
therefore only modify those pixels that are considered during
scaling. The resulting sparse noise is visually imperceptible, yet
controls the entire output of the scaling process
(see~\autoref{fig:scaling_attack_rootcause}).  In this way, both
attack objectives, $\scalefunc(\attimg) \sim \tarimg$ and
$\attimg \sim \srcimg$, are fulfilled.  We build our detection
defenses on this understanding of scaling attacks.

\paragraph{Prevention Defenses}
We also recap defenses that prevent image-scaling
attacks.
Prior work has studied two concepts: First, we can apply a robust
scaling algorithm that considers all pixels with equal contribution.
This directly addresses the root cause of the attack, since no pixels
are ignored anymore.  A possible scaling algorithm is \emph{area
  scaling} that is implemented in several image processing libraries.
Second, the defender can use an image-reconstruction method to repair
the modified pixels from their neighborhood. To circumvent this
defense, an adversary would have to modify the neighborhood as well,
which leads to visible modifications. We refer the reader to the paper
by \citet{QuiKleArp20} which examines both concepts and their
robustness.

While prevention mechanisms do not interfere with the typical
machine-learning workflow, they have a clear disadvantage: The
mechanisms cannot be applied to find out \emph{that an attack is going
  on}, that is, the data is manipulated. However, this can be
necessary to ensure that we can trust a specific dataset, for
instance, if we use publicly available images or
release our image database. Hence, there is a need to study
effective \emph{detection} methods as complementary approach
to existing prevention concepts.

\section{Detection Systematization}
\label{sec:detection-methods}

The detection of image-scaling attacks has received little focus so far.  While some heuristics have been proposed~\citep{KimAbuGao+21, XiaCheShe+19}, we still lack a general understanding on how the attacks can be effectively characterized and detected. To fill this gap, we systematize existing work by identifying two general \emph{detection paradigms}.  For each paradigm, we present the basic principle of detection and introduce own realizations that alleviate shortcomings of existing heuristics.  \autoref{tab:overview_detection_methods} shows an overview of all considered detection methods.

\newcommand{\msessim}{$\lbrace$MSE, SSIM$\rbrace$\xspace}
\newcommand{\psnrssim}{$\lbrace$PSNR, SSIM$\rbrace$\xspace}
\begin{table}
	\caption{Overview of detection methods in this paper.}
	\label{tab:overview_detection_methods}
	\vspace{-0.6em}
	\begin{tabularx}{\columnwidth}{
			llll
		}
		\toprule
		\multicolumn{2}{c}{Paradigm} & Method & 
		Options \\
		\midrule
		\multirow{4}{*}{\shortstack[l]{\rotatebox[origin=c]{90}{Frequency}}}
		& 
		& \thisw Peak Spectrum & \\
		&& \thisw Peak Distance & \\
		&& \thiswn CSP~\cite{KimAbuGao+21} & \\
		&& \thisw CSP--improved & \\
		\cmidrule{2-4}
		\multirow{9}{*}{\shortstack[l]{\rotatebox[origin=c]{90}{Spatial}}}
		& 
		\multirow{5}{*}{\shortstack[l]{\rotatebox[origin=c]{90}{Adversarial}}}
		& \thisw Down \& Upscaling & PSNR \\ 
		&& \thiswn Down \& Upscaling & $\lbrace$Histogram, 
		Color-scattering$\rbrace$~\cite{XiaCheShe+19} \\
		&& \thiswn Down \& Upscaling & \msessim~\cite{KimAbuGao+21} \\
		&& \thiswn Maximum filter & \msessim~\cite{KimAbuGao+21} \\
		&& \thiswn Minimum filter & \msessim~\cite{KimAbuGao+21} 
		\\		
		\cmidrule{3-4}
		& 
		\multirow{3}{*}{\shortstack[l]{\rotatebox[origin=c]{90}{Clean}}}
		& \thisw Clean Filter & $\lbrace$Median, 
		Random$\rbrace$+\psnrssim \\
		&& \thisw Patch-Clean Filter & \\
		&& \thisw Targeted Patch-Clean Filter & \\
		\bottomrule
	\end{tabularx}
\begin{tablenotes}
	\centering
	\footnotesize
	\item \thisws=Proposed in this paper.
	\item $\lbrace . \rbrace$ denotes a set of possible 
	options for a respective method.
\end{tablenotes}
\end{table}

\subsection{Paradigm: Frequency Analysis}
\begin{figure}[t]
	\centering
	\subcaptionbox{Spectrum of~\srcimg
		\label{fig:frequency_detection_example_a}}
	[.32\columnwidth]
	{\includegraphics[scale=0.0682]{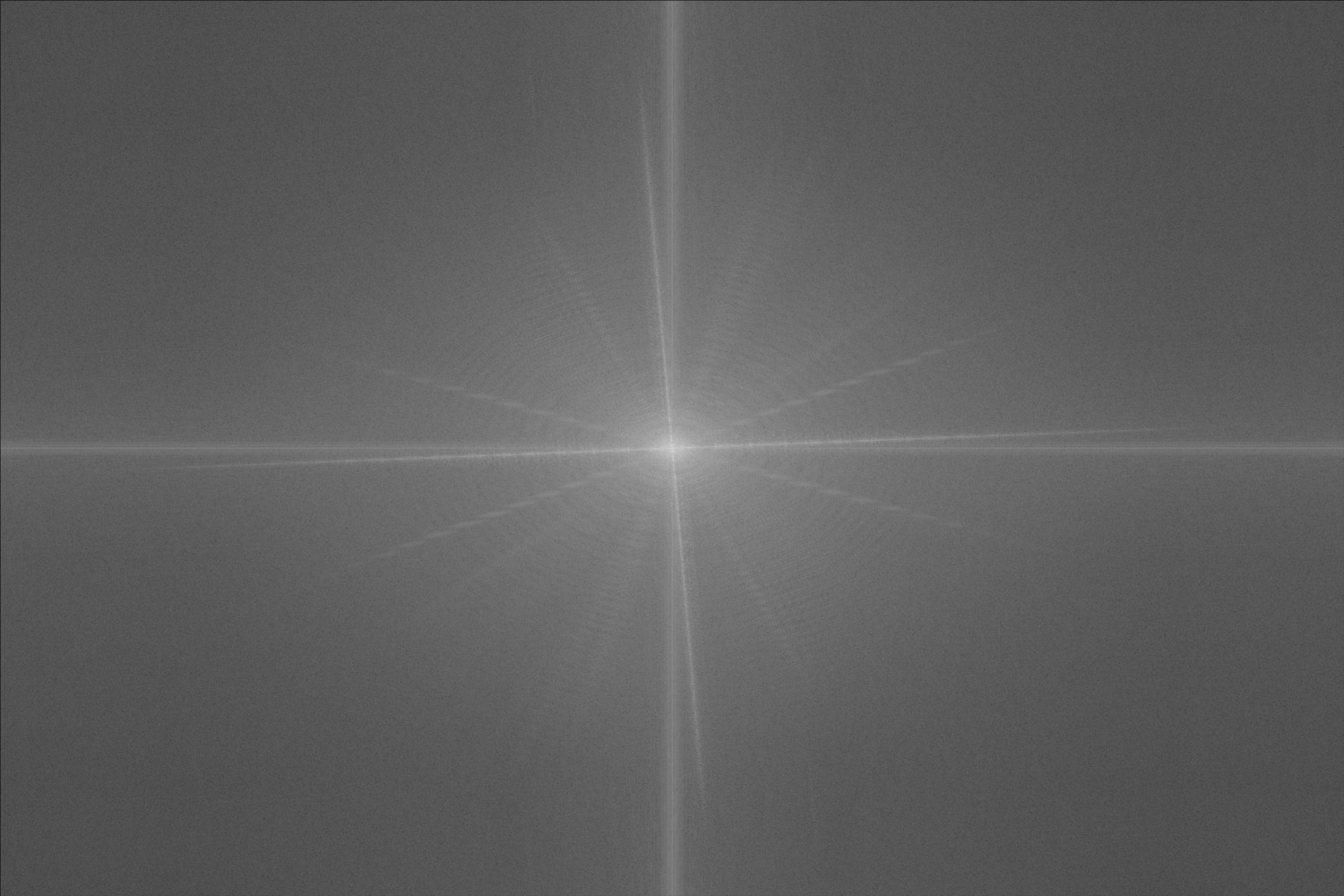}}
	\subcaptionbox{Spectrum of~\attimg
		\label{fig:frequency_detection_example_b}}
	[.32\columnwidth]
	{\includegraphics[scale=0.0682]{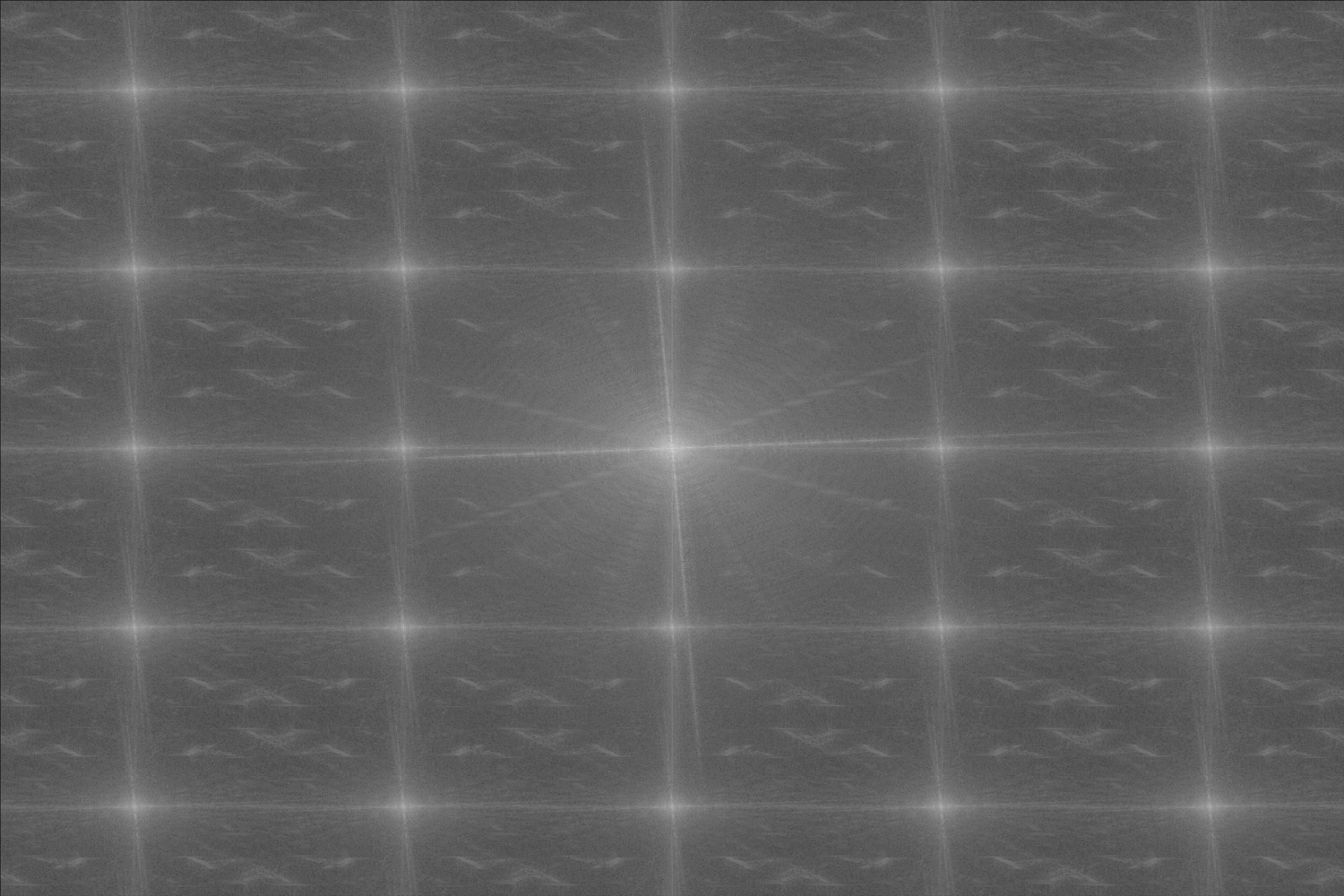}}
	\subcaptionbox{Marked peaks in \attimg
		\label{fig:frequency_detection_example_c}}
	[.32\columnwidth]
	{\includegraphics[scale=0.0682]{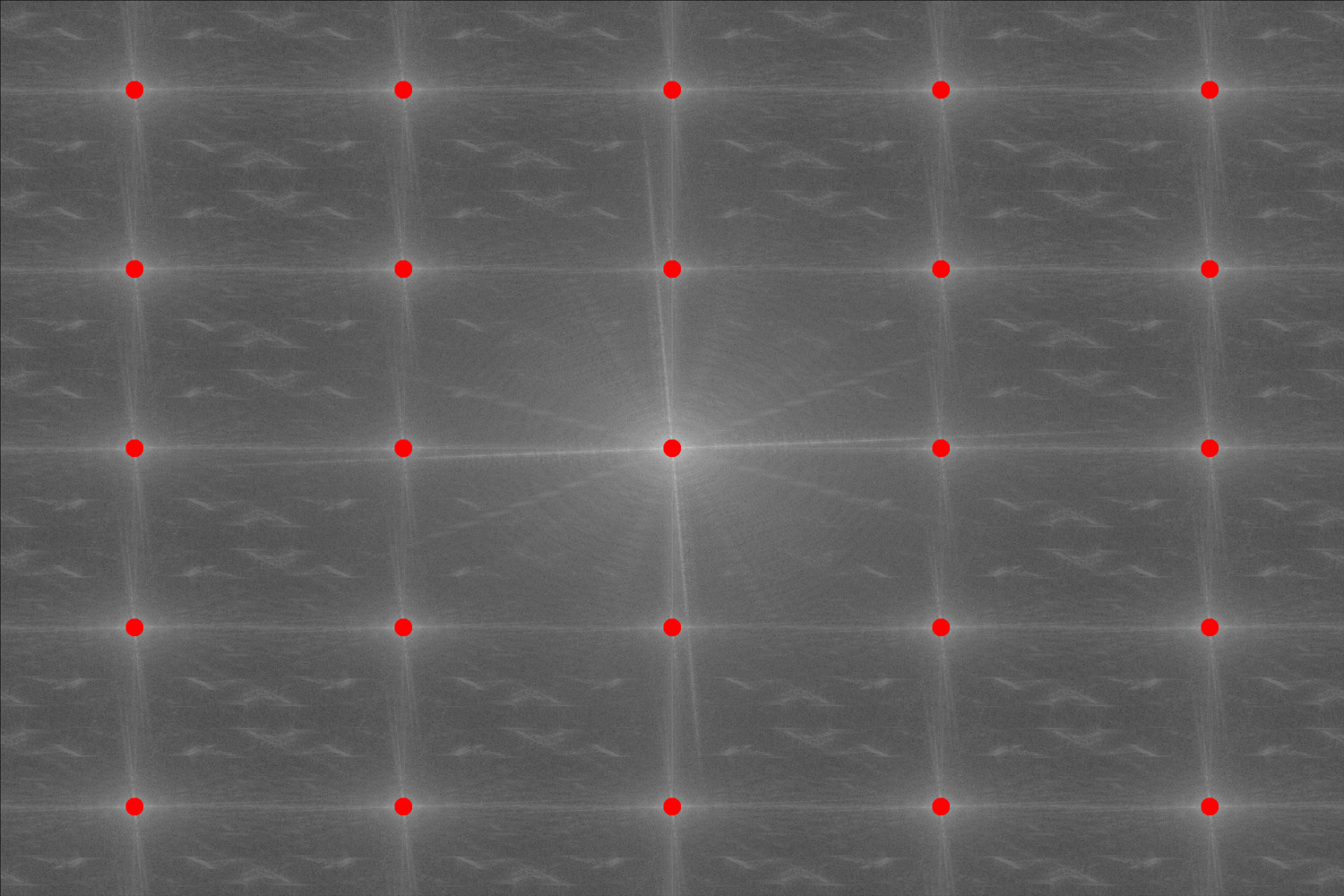}}
	\vspace{-0.55cm}
	\caption{Frequency analysis of the example in
		\autoref{fig:scaling_attack_example}. Plot (a) and (b) show 
		the 
		frequency spectrum of the source image and attack image, 
		respectively. Plot (c) shows the marked peaks with 
		\autoref{eq:peak-function-centered}.
	}
	\label{fig:frequency_detection_example}
\end{figure}

In this paradigm, the detection builds on analyzing the frequency
spectrum of images. This is a common procedure in computer vision and
multimedia security~\cite{SenMem13}. In this frequency representation,
periodical patterns become evident that are not detectable in the
spatial image domain (pixel domain)~\cite{KirBoe08}.  
In general, a frequency
spectrum is an equivalent representation of an image that describes
the pixels by a sum of waves oscillating at different
frequencies~\cite{Smi97}. In this work, we use the 2D discrete Fourier
transform (DFT) and work on the centered log-scaled magnitude spectrum
(as visible e.g.\ in \autoref{fig:frequency_detection_example_a}).
Intuitively, each coefficient in this magnitude spectrum shows the
impact of a particular frequency for the image. The log-scaling
emphasizes smaller values. The middle of the magnitude spectrum
corresponds to low frequencies while higher frequencies are located
towards the corners.

As the root-cause analysis of image-scaling attacks in \autoref{sec:background} highlights, the
adversary injects pixels from the target image~\tarimg into \srcimg in
a periodic distance. This periodicity stems from the 
sampling process
of image scaling and provides a strong indicator of image-scaling
attacks. Hence, a modified image has unique, periodic peaks in its
frequency spectrum. As an example, let us
analyze the running example in
\autoref{fig:scaling_attack_example}. The modification is not visible
in the spatial domain. However, if we analyze the attack image in the
frequency domain---as visible in
\autoref{fig:frequency_detection_example_b}---one can clearly see 
unusual frequency peaks.

\paragraph{Proposed Detection}
For our analysis, we take inspiration from multimedia
forensics~\cite{SenMem13, CheHsu11}.  We start by noting that the
defender exactly knows the potentially modified pixels in the spatial
domain.  This allows an exact computation of the expected peak
locations in the frequency spectrum.  As a result, we gain the
advantage of differentiating adversarial peaks created by scaling
attacks from benign peaks that images can naturally have at other
locations in the spectrum, reducing the chances to detect benign peaks
accidentally.

In particular, our detection approach proceeds as follows:
Let $(m,n)$ be the height and width of the source image and $(m', 
n')$ the height and width of the scaled output image. The vertical 
scaling ratio is given as $\Tm = \tfrac{m}{m'}$ while the horizontal 
scaling ratio is $\Tn = \tfrac{n}{n'}$. The constants $c_m$ and $c_n$ 
are the index of the spectrum's middle. In the case of a scaling 
attack, the following binary function \mbox{$\Gamma \in \lbrace 0, 1 
\rbrace ^{m \times n}$} shows at which frequency coefficient a peak 
occurs:
\begin{align}
	\Gamma(u,v) =
	\begin{cases}
		1 &
		(u, v) = (c_m + k_1 \cdot m', c_n + k_2 \cdot n') \\
		0 & \text{otherwise}.
	\end{cases}
	\label{eq:peak-function-centered}
	\\
	\text{with} 
	-\tfrac{\Tm}{2} \leqslant k_1 \leqslant \tfrac{\Tm}{2}, \; 
	-\tfrac{\Tn}{2} \leqslant k_2 \leqslant \tfrac{\Tn}{2}, \;
	k_1, k_2 \in \mathbb{N}. \nonumber
\end{align}
In other words, we expect to observe peaks around 
each \mbox{$k_1 m'$-th} and $k_2 n'$-th 
position of the frequency spectrum if an image is manipulated by an 
image-scaling attack.
In \appref{sec:appendix-frequency-analysis}, we derive 
\autoref{eq:peak-function-centered}. 
To provide some intuition, 
\autoref{fig:frequency_detection_example_c} marks the expected 
peaks on the frequency spectrum of our running example by using 
\autoref{eq:peak-function-centered}. One can see that the observed 
and expected peaks match exactly.
Equipped with the ability to predict the expected peaks, we propose 
two detection strategies.

\begin{figure}
	\centering
	\captionsetup[subfigure]{%
		aboveskip=-0pt,belowskip=-0pt}
	\subcaptionbox{Peak-Spectrum Analysis
		\label{fig:frequency_detection_approaches_a}}
	[.9\columnwidth]
	{\includegraphics{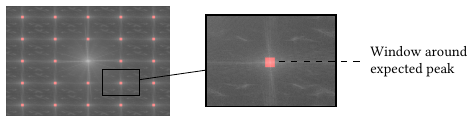}}
	\subcaptionbox{Peak-Distance Analysis
		\label{fig:frequency_detection_approaches_b}}
	[.9\columnwidth]
	{\includegraphics{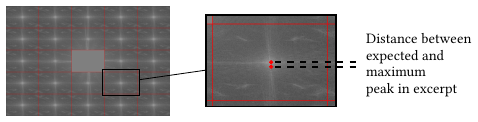}}
	\vspace{-0.2cm}
	\caption{Frequency-based detection approaches. Plot~(a): Windows 
		around expected peaks are extracted and their frequencies are
		compared in relation to the overall spectrum. Plot (b):
		Spectrum is divided into excerpts and the distance between 
		expected 
		and maximum peak for each excerpt is computed.
	}
	\label{fig:frequency_detection_approaches}
\end{figure}

\vspace{0.3em}
\minipara{Peak Spectrum} We extract   
  the frequencies in a square window %
  centered around all expected peak locations. The window's half 
  length is $\w$.
  We omit the center $(c_m, c_n)$ being present
  in any spectrum.  \autoref{fig:frequency_detection_approaches_a}
  illustrates the resulting windows.  
  Next, we average all window values and calculate the
  percentile rank of this value relative to the whole frequency
  spectrum. This sets the peak frequencies into relation
  to the entire frequency spectrum. In case of an attack, the peak 
  frequencies outshine the entire spectrum, 
  so that attack images get higher percentile ranks than benign 
  images.

\vspace{0.3em}
\minipara{Peak Distance}  We divide the spectrum
  into excerpts for each peak as
  \autoref{fig:frequency_detection_approaches_b} visualizes. We
  discard the center of the spectrum again. For each excerpt, we
  extract the maximum peak and calculate the distance between this
  peak and the expected peak. We then average all measured distances.
  In case of an attack, this average is expected to be small.

\paragraph{Previous Approaches}
\citet{KimAbuGao+21} propose a frequency analysis, named 
\emph{CSP}, where the underlying idea is to count peaks at arbitrary 
positions in the spectrum. Based on their evaluation, they assume an 
attack if the spectrum contains more than one peak. Our evaluation 
shows that this approach is ineffective. Benign images can naturally 
have peaks, too. We show two such examples from ImageNet in
\appref{sec:appendix-peaks-benign-images}.
To test if the method works without the fixed threshold of one, 
we evaluate an own adjustment, named
\emph{CSP-improved}, where we derive the threshold from the data. 
Still, this adjustment cannot compete with our methods. 
The defender should rather
use the advantage of having precise knowledge about the 
expected peak locations---as our proposed methods do.

\subsection{Paradigm: Spatial Analysis}

In this paradigm, the analysis is done in the spatial domain.  As
scaling algorithms and scaling attacks operate here, this domain
naturally provides the advantage of knowing which pixels are
considered by scaling algorithms and hence which are possibly
modified.  We identify two variants of this paradigm:
A defense leverages either the \emph{adversarial} 
modification or the \emph{clean} pixels in~\attimg for 
detection. In the following, we examine each group in more detail. 

\subsubsection{Adversarial-Signal Driven}
The concept here is to amplify the sparse, adversarial modifications
in \attimg and then to compare the amplified image $\attimg'$ with 
\attimg. We have two sub-groups here.

\paragraph{Down-and-Upscaling}
We can exploit that downscaling and upscaling form antagonists in the
spatial domain. That is, the scaling of \attimg leads to an output
\outimg which corresponds to \tarimg. By upscaling \outimg back to the
original size, $\attimg' = \upscalefunc(\downscalefunc(\attimg))$, we
can compare this version with \attimg. The upscaling strengthens the
signal embedded by the adversary and renders it detectable through
comparison with the original image. Note that the comparison
function needs to account for this setup, as we describe later.

In principle, we do not need to upscale the image and could directly
compare \attimg with \outimg. In our preliminary experiments, however,
we found that we can achieve better results if we upscale \outimg to
the resolution of \attimg and conduct the comparison on the same
dimensions. Moreover, some image comparison methods require the same
size for the inputs.

\minipara{Proposed Detection}
A variety of methods exists for comparing images that would be
applicable here. 
We find that a simple
PSNR computation is particularly effective. Peak signal-to-noise ratio
(PSNR) is a widely used pixel-based comparison method that returns the
normalized mean-squared error between two images. Formally, it is
defined as
\begin{align}
	\mathrm{PSNR}(\attimg, \attimg') = 20 \log_{10}
	(I) - 10 \log_{10}(\mathrm{MSE}(\attimg, \attimg')) .
	\label{eq:psnr}
\end{align}
The constant $I$ is the maximum of the pixel range, such as 255 for
8-bit images. The higher the PSNR value is, the more two images match.
PSNR can be computed efficiently and thus provides a perfect basis for
designing a detection method using down- and upscaling.

\vspace{0.3em}
\minipara{Previous Approaches}
\citet{XiaCheShe+19} propose two alternative comparison options. 
First, the \emph{intensity histograms} of \attimg and of $\attimg'$ 
are extracted and compared.
Second, in the \emph{color-scattering} method, the 
average distance to the image center over all pixels with the same 
intensity is computed. By doing this for each intensity value in 
\attimg and $\attimg'$, respectively, we obtain two vectors that can 
be compared. 
Moreover, \citet{KimAbuGao+21} propose two measures
that directly
compare \attimg and $\attimg'$: the mean squared error (MSE) and the
SSIM index. 
While the first is simply comparing \attimg and $\attimg'$
pixel-wise, the latter aggregates information about the luminance,
contrast, and structure between \attimg and 
$\attimg'$~\cite{HorZio10}.

In our evaluation, the simple PSNR and SSIM outperform 
the other measures.
Unlike suggested by prior work~\cite{KimAbuGao+21}, MSE is not
superior to PSNR. We assume that this conclusion is the result of 
a subtle implementation mistake in the evaluation metric where an 
integer overflow occurs. In fact, the PSNR and MSE are directly 
related to each other (see \autoref{eq:psnr}). Although MSE and 
PSNR have the same detection performance, the PSNR score is 
easier to interpret. It typically lies between 0~dB and 60~dB in our 
evaluations.
Moreover, in terms of security, a pixel-wise measure such as the PSNR 
is particularly robust against adaptive attacks compared to
aggregation-based measures such as the histogram or
color-scattering. No information are lost that could be
exploited~\cite{QuiRie20}.

\paragraph{Amplifying Filter}
As alternative to down-and-upscaling,
\citet{KimAbuGao+21} propose two methods, a minimum filter and a 
maximum filter, to strengthen the adversarial signal.
Each filter is applied on
\attimg by iterating over the image and replacing each pixel by the
minimum/maximum of its neighborhood. The outcome~$\attimg'$ is 
compared with \attimg using MSE or SSIM.
This filtering strategy aims at amplifying the periodic modification. 
However, as the CSP approach, this defense overlooks 
the attack's root cause and misses the advantage of knowing the 
pixels used by scaling algorithms. A targeted filtering
is possible which motivates our next group of defenses.

\subsubsection{Clean-Signal Driven}
Another concept of the spatial paradigm 
is to leverage the clean pixels that a scaling attack 
needs to leave to keep the attack imperceptible.
We can use these pixels to clean \attimg and to compare this result 
with the initial version.
This principle has not been studied before. %
It allows us to design specialized methods for the global and local 
attack scenario, which are presented in the following.

\paragraph{Proposed Global Detection}
To obtain a cleansed version of~\attimg, we adopt the idea of
reconstructing the modified pixels using \emph{prevention 
filters}~\cite{QuiKleArp20}.  In particular,
we design two methods that are based on a \emph{selective median 
filter} and a \emph{selective random filter}, respectively. Both 
filters repair each pixel considered by a scaling algorithm by using 
the neighborhood of the particular pixel.
In terms of security, the reconstruction methods have the advantage 
of addressing the root cause of scaling attacks. They also show a 
strong robustness against adaptive attacks, since adversaries have to 
modify the neighborhood as well to bypass both filters, making the 
attack clearly visible~\cite{QuiKleArp20}. This security robustness 
also transfers to our detection defense.

For detection, we propose the following simple \emph{clean filter} 
approach: 
Equipped with a prevention filter~$\mathcal{V}$, we can get a cleaned 
version $\attimg' = \mathcal{V}(\attimg)$ which is compared with the 
initial input \attimg. We evaluate the PSNR and the SSIM measure to 
compare $\attimg$ and $\attimg'$.

Note that we also tested to compare $\outimg$ and $\outimg'$ where
$\outimg'$ is the downscaled version of $\attimg'$. Hence, \outimg 
shows the new adversarial content, while $\outimg'$ shows the 
original content. The detection results correspond to the 
clean-filter method and are thus omitted.

\paragraph{Proposed Local Detection}
A direct application of the previous approach to the local attack
scenario is not promising. As \attimg and $\attimg'$ are here similar 
for the majority of pixels, a global comparison cannot sufficiently 
capture the difference. Instead, we propose to divide the images into
\emph{patches} and compare each patch individually. We propose two 
variants.

\vspace{0.3em}
\minipara{Patch-Clean Filter}
We create $\outimg$ and~$\outimg'$ and
divide them into $L$ patches, 
respectively. 
We compute the PSNR between each pair of patch, 
\mbox{$v_i = 
\mathrm{PSNR}(\outimg_{i}, \outimg'_{i}) \; \forall i = 1,
\dots, L$}. The final detection score is given as
$
	\vert \mathrm{mean}(\lbrace v_i \rbrace) - \mathrm{min}(\lbrace 
	v_i \rbrace) \vert.
$
\appref{sec:appendix-patch-local-detection} presents the method in 
more detail. 
Note that we work on downscaled images here. We find that the smaller 
image size reduces the number of possible regions which improves the 
detection rate.

\vspace{0.3em}
\minipara{Targeted Patch-Clean Filter}
As alternative, we analyze the unscaled images. However, even 
with patches, the scaling pixels---and thus modified pixels---are in 
the minority, making it difficult to detect a difference with and 
without scaling attack. Hence, we propose a more targeted variant and 
only examine the scaling pixels in each patch. 
Let $L$ denote the number of patches, $\attimg_i$ \& 
$\attimg'_i$ the respective patches, and 
$\Psi$ the scaling-pixel selection, we compute
\begin{align}
u_i = \vert\Psi(\attimg'_i) - \Psi(\attimg_i)\vert \quad \forall i = 
1,\dots, L \; .
\end{align}
We get the $q$-th quantile of each vector $u_i$, denoted as
\mbox{$\mathrm{q}(u_i)$}, and calculate 
$
\vert \mathrm{max}(\lbrace \mathrm{q}(u_i) \rbrace) - 
\mathrm{mean}(\lbrace \mathrm{q}(u_i) \rbrace) \vert
$
as detection score. This comparison allows us to identify an unusual 
difference in a local area.
Note that the choice of $q$ is a hyperparameter of the method.

\vspace{0.3em}
\minipara{Remark}
We also tested the down-and-upscaling concept with patches. However, 
$\attimg' = \upscalefunc(\downscalefunc(\attimg))$ contains a 
stronger noise signal due to down-and-upscaling. While 
this is tolerable in the global scenario, it distorts the comparison 
with small patches. 

\section{Evaluation}
\label{sec:evaluation}
We proceed with an empirical evaluation of the different detection
methods. To obtain a comprehensive view, we
study both non-adaptive and adaptive adversaries. In this section, we
consider the non-adaptive case with regular attacks. Then, in
\autoref{sec:adaptiveattack}, we investigate the best approaches 
against an adaptive adversary.

\subsection{Evaluation Setup}
\label{subsec:evaluation-setup}
Our evaluation follows the common design of experiments on image-scaling attacks~\cite{XiaCheShe+19, QuiKleArp20}.
We evaluate the detection of attacks against popular scaling
algorithms that are vulnerable to scaling 
attacks~\mbox{\cite{QuiKleArp20}}. In particular, we consider 
nearest-neighbor, bilinear, and
bicubic scaling from the libraries OpenCV and tf.image
(TensorFlow), and nearest-neighbor scaling from the 
library Pillow. We omit Lanczos scaling which is
comparable to bicubic scaling~\mbox{\cite{QuiKleArp20}}.

For the attacks, we adopt the evaluation setup by \citet{QuiKleArp20},
where the source and target for an attack are randomly drawn from a
collection of images. As dataset for this sampling, we use photos from
ImageNet~\cite{RusDenSu+15}.  Compared to other datasets from computer
vision, like CIFAR or CelebA, ImageNet contains significantly larger
images, which is a key requirement for constructing successful scaling
attacks in practice. Moreover, the dataset is very diverse, covering
various image sizes and contents, such as faces, animals, persons,
objects, and landscapes.

As learning model, we use a pre-trained VGG19 model~\citep{SimZis14} 
which is a standard benchmark in computer vision. The target size for 
scaling is thus $224 \times 224 \times 3$ pixels.
Note that we just use this one architecture in our experiments, since
scaling attacks do not depend on the learning model's architecture. 
They change the input to the model. Only the input size of 
the model is relevant, so that we make sure to use varying scaling 
ratios as described later.
Finally, we consider a global and local modification scenario. Both
require a slightly different setup that we present in the following.

\paragraph{Global Modification}
To obtain attack images in this scenario, we randomly sample
images from ImageNet and create 1,000 source--target image pairs. We 
ensure that the pairs have varying scaling ratios to avoid 
artifacts that may 
arise from a fixed ratio. We check that each target is unrelated to 
its source image by requiring different classes and predictions for 
each pair.
After conducting the attacks, we keep only those images that are
successful regarding both attack objectives, \ie
$\scalefunc(\attimg) \sim \tarimg$ and $\attimg \sim \srcimg$.  To
this end, we use the same methodology as~\citet{QuiKleArp20}.
The number of successful images varies across each combination of
scaling algorithm and library, so that we choose the highest possible
number of images across all setups.  This leads to 585 attack images
for each combination of scaling algorithm and library. As unmodified
reference set, we additionally select 585 further images from
ImageNet. They are used to evaluate the detection of benign inputs.

\paragraph{Local Modification}
Here, we implement the \emph{BadNets} backdoor
attack~\citep{GuDolGar17} as a representative form for local triggers
that are also used in recent works~\mbox{\cite[e.g.,][]{SalWenBac22}}.
We use the same 585 source images as in the previous scenario. Yet, we
scale each source image to a size of $224 \times 224 \times 3$ pixels
and add a small, bounded backdoor pattern to create its respective
target image (see~\autoref{fig:appendix_eval_backdoor_examples} in the
Appendix).  Finally, we conduct scaling
attacks on these image pairs.  As reference set, we use the same
585 benign images as before.
We also verify that the so-created backdoors are effective (see 
\appref{sec:appendix-backdoor-setup}).

\paragraph{Evaluation Measures}
To evaluate the detection performance of the considered detection
methods, we equally split each attack and reference dataset into a
training and test partition, respectively.  The training set is used
to calibrate a threshold for each detection method so that the false
positive rate is 1\%. This threshold is then used to evaluate the
detection performance on the test set.
For some detection methods, we need to decide on specific
hyperparameters, such as the window width in the peak-spectrum
analysis. For this calibration, we split the training set further and
create a validation set. This set is used to find the optimal
parameters in a grid search. We
instantiate the detection methods with the best parameters (see 
\appref{sec:appendix-hyperparams}) and report
their performance on the test set.

\subsection{Global Modification Scenario}
\label{subsec:eval-global-scenario}
We start with the detection when source
and target image are arbitrary. This is the most severe setting as an
adversary can freely choose the target class of the prediction.

\paragraph{Results}
\autoref{tab:eval_full_image_modification} shows the performance for
all detection methods, sorted in descending order with respect to the
average accuracy. Both paradigms allow detecting scaling attacks if
the entire scaled image is modified. A perfect detection is possible
with our proposed peak-distance frequency analysis. However, not all
methods are effective, such as the previously proposed CSP approach
with an accuracy of 50\%.

\begin{table}
	\centering
	\begin{tabularx}{\columnwidth}{llRRR}
\toprule
                     \thiswn Method &               Option &  AvgAcc &  StdAcc &  AvgFPR \\
\midrule
               \thisw Peak Distance &                      & 100.00 &  00.00 &  00.00 \\
               \thisw Peak Spectrum &                      &  99.90 &  00.26 &  00.00 \\
                \thisw Clean Filter &  Median filter, SSIM &  99.80 &  00.21 &  00.05 \\
                  \thiswn Down \& Up &            Histogram &  98.85 &  01.09 &  00.49 \\
                   \thisw Down \& Up &                 PSNR &  98.54 &  01.48 &  00.49 \\
                  \thiswn Down \& Up &                  MSE &  98.54 &  01.48 &  00.49 \\
                \thisw Clean Filter &  Random filter, SSIM &  98.54 &  01.51 &  00.63 \\
 \thisw Targeted Patch-Clean Filter &                      &  96.54 &  04.16 &  00.88 \\
                \thisw Clean Filter &  Median filter, PSNR &  94.30 &  06.06 &  01.07 \\
             \thiswn Maximum Filter &                 SSIM &  87.03 &  03.53 &  01.80 \\
             \thiswn Minimum Filter &                 SSIM &  85.15 &  04.11 &  01.66 \\
                \thisw Clean Filter &  Random filter, PSNR &  81.52 &  08.59 &  01.17 \\
                         \thisw CSP &             Improved &  76.72 &  06.19 &  01.32 \\
                  \thiswn Down \& Up &                 SSIM &  76.21 &  12.95 &  01.46 \\
             \thiswn Minimum Filter &                  MSE &  64.65 &  03.15 &  01.61 \\
             \thiswn Maximum Filter &                  MSE &  60.63 &  01.39 &  01.66 \\
                  \thiswn Down \& Up &     Color-scattering &  55.88 &  03.70 &  00.83 \\
                        \thiswn CSP &             Original &  50.00 &  00.00 &  00.00 \\
          \thisw Patch-Clean Filter &                      &  49.95 &  00.08 &  00.10 \\
\bottomrule
\end{tabularx}

	\vspace{0.15em}
	\caption{Detection performance [\%] in global scenario 
		in terms of 
		accuracy (average + standard dev.) and false positives 
		(average)
		over all scaling algorithms and libraries.
		$\,$ \thisws Proposed in this paper.
	}
	\label{tab:eval_full_image_modification}
\end{table}

\paragraph{Analysis}
A closer look on this experiment provides important insights.  First,
our proposed frequency methods---exploiting the known peak
locations---outperform all other detection methods. Their accuracy is
100\% and 99.90\%.  On the other hand, the CSP
approach~\cite{KimAbuGao+21} that also analyzes the frequency spectrum
is comparable to random guessing.  
Second, the choice of the comparison function in the spatial paradigm
is important. SSIM is preferable for the clean, minimum, and maximum 
filter. 
PSNR provides a higher detection 
accuracy with the down-and-upscaling approach.
There is no difference between MSE and PSNR in our experiments. 
Third, patch-based defenses designed for the 
local scenario are partly applicable in the global case. 
The targeted patch-clean filter has a detection rate of 96.54\%.

\paragraph{Ensemble of Detection Methods}
Next, we combine multiple methods as 
ensemble to increase the diversity of detection patterns. We test 
three ensembles: We use the best method from each paradigm, and we 
use the $K=\lbrace3,4\rbrace$ best methods (irrespective of the 
paradigm). Note that we choose the methods based on the results on 
the training dataset and not 
\autoref{tab:eval_full_image_modification}.

The first ensemble consists of peak distance and the 
clean filter (with Median, SSIM). 
The ensemble with $K=3$ consists of the first three entries in 
\autoref{tab:eval_full_image_modification}. The ensemble with $K=4$ 
uses Down \& Up with PSNR as 4th method in addition.
We use two voting strategies: We report an attack 
if the \textit{majority} of methods or if at least one 
method flags an input, that is, \textit{one winner takes all}. 

\autoref{tab:eval_full_image_modification_ensemble} shows the 
performance. The K-best ensemble with \mbox{$K=4$} and majority 
voting achieves an 
accuracy of 100\% and a false-positive rate of 0\%. In 
terms of security, an ensemble thus allows a perfect detection 
rate while increasing the difficulty for an adaptive attack, as 
different paradigms and methods have to be circumvented.

\begin{table}[t]
\centering
\setlength\tabcolsep{3pt} %
\renewcommand{\arraystretch}{1.2}%
\begin{tabularx}{1\linewidth}{lRRRRRR}
\toprule
& \multicolumn{3}{c}{Majority} & \multicolumn{3}{c}{One Winner Takes 
All} \\
 \cmidrule(lr){2-4} \cmidrule(lr){5-7}
Ensemble & Acc & TPR & FPR & Acc & TPR & FPR \\
\midrule
Best Per Paradigm & 99.98 $\pm$~0.06& 100.00 $\pm$~0.00 & 0.05 
$\pm$~0.13 & --- & --- & --- \\
$K$ Best ($K$=3) & 99.98 $\pm$~0.06& 99.95 $\pm$~0.13 & 0.00 
$\pm$~0.00 & 99.98 $\pm$~0.06& 100.00 $\pm$~0.00& 0.05 $\pm$~0.13\\
$K$ Best ($K$=4) & 100.00 $\pm$~0.00& 100.00 $\pm$~0.00 & 0.00 
$\pm$~0.00 & 99.73 $\pm$~0.17& 100.00 $\pm$~0.00& 0.54 $\pm$~0.33\\
\bottomrule
\end{tabularx}
\vspace{0.15em}
\caption{Ensemble in global scenario. Each 
cell shows the average $\pm$ standard deviation over all scaling 
libraries and algorithms. With two paradigms, both 
voting methods for ``best per paradigm'' have
identical results, so that we omit the 2nd voting method in this 
case.}
\label{tab:eval_full_image_modification_ensemble}
\vspace{-0.45cm}
\end{table}

\paragraph{Overlay Scenario}
In addition, we study the variation where a scaling attack is used 
to embed only a low-opacity version of $\tarimg$ into the downscaled 
output. More specifically, the novel target image is given 
as 
$ %
\tarimg' = \alpha \cdot \tarimg + (1-\alpha) \cdot 
\scalefunc(\srcimg) .
$ %
The parameter $\alpha$ denotes the blending factor. In our 
experiments, we set $\alpha=0.3$ to test a challenging case with only 
a very small embedding of \tarimg.

\autoref{tab:eval_global_image_overlay_0_3} shows the performance. 
Our frequency approaches still provide a 
detection rate close to 100\%. Even with a very low blending factor, 
periodic peaks are inevitably created in the frequency spectrum. On 
the contrary, the spatial paradigm is more affected by the overlay 
scenario. A pixel-based comparison is more difficult, as 
$\tarimg$ is embedded more weakly.

\paragraph{Comparing Scaling Algorithms and Libraries}
We have presented aggregated results over all scaling 
algorithms and libraries so far. In 
\appref{sec:appendix-evaluation-scalesetups}, 
we analyze the individual detection per scaling algorithm and 
library. We find no significant difference between the libraries. 
Moreover, the frequency methods and the leading methods in the 
spatial paradigm do not depend on the scaling algorithm.

\paragraph{Summary}
The effective detection of image-scaling attacks is possible with both
paradigms. The frequency paradigm allows for a perfect detection rate
without any false positives in our experiments.  Even in the overlay
scenario, it enables spotting all attacks with high accuracy.  We
conclude that scaling attacks can be efficiently detected if an
adversary performs a global modification.

\subsection{Local Modification Scenario}
\label{subsec:eval-local-scenario}
In our next experiment, we test the detection performance in the
challenging scenario where an adversary applies an image-scaling
attack only to a small area of an image.

\paragraph{Results}
\autoref{tab:eval_limited_modification} shows the results for all
detection methods. Only our proposed frequency methods can effectively
detect local scaling attacks with an average accuracy of 80.98\% and
89.81\%.  Our patch-based approaches achieve an acceptable detection
rate of 76.38\% and 75.72\%. The other methods are close to random 
guessing.

\begin{table}
	\centering
	\begin{tabularx}{\columnwidth}{llRR}
\toprule
       \thiswn Method &               Option &  AvgAcc &  StdAcc \\
\midrule
 \thisw Peak Distance &                      &  99.73 &  00.24 \\
 \thisw Peak Spectrum &                      &  99.71 &  00.13 \\
  \thisw Clean Filter &  Median filter, SSIM &  93.47 &  05.12 \\
    \thiswn Down \& Up &            Histogram &  92.54 &  06.91 \\
  \thisw Clean Filter &  Random filter, SSIM &  78.30 &  08.99 \\
\bottomrule
\end{tabularx}

	\vspace{0.15em}
	\caption{Detection accuracy in overlay scenario. Only the 
		effective approaches with AvgAcc > 60\% are shown.}
	\label{tab:eval_global_image_overlay_0_3}
\end{table}

\begin{table}
	\centering
	\begin{tabularx}{\columnwidth}{llRRR}
\toprule
                     \thiswn Method &               Option &  AvgAcc &  StdAcc &  AvgFPR \\
\midrule
               \thisw Peak Spectrum &                      &  89.81 &  04.75 &  01.07 \\
               \thisw Peak Distance &                      &  80.98 &  02.83 &  01.17 \\
 \thisw Targeted Patch-Clean Filter &                      &  76.38 &  12.35 &  00.93 \\
          \thisw Patch-Clean Filter &                      &  75.72 &  04.91 &  01.71 \\
                  \thiswn Down \& Up &     Color-scattering &  51.56 &  02.08 &  00.83 \\
                  \thiswn Down \& Up &            Histogram &  50.44 &  01.13 &  00.73 \\
                \thisw Clean Filter &  Median filter, SSIM &  50.15 &  00.18 &  00.59 \\
                \thisw Clean Filter &  Random filter, SSIM &  50.02 &  00.06 &  00.49 \\
                        \thiswn CSP &             Original &  50.00 &  00.00 &  00.00 \\
             \thiswn Maximum Filter &                 SSIM &  49.85 &  00.06 &  01.02 \\
             \thiswn Minimum Filter &                 SSIM &  49.80 &  00.51 &  01.46 \\
                \thisw Clean Filter &  Median filter, PSNR &  49.78 &  00.16 &  00.98 \\
                  \thiswn Down \& Up &                  MSE &  49.68 &  00.35 &  00.83 \\
                   \thisw Down \& Up &                 PSNR &  49.68 &  00.35 &  00.83 \\
                \thisw Clean Filter &  Random filter, PSNR &  49.68 &  00.12 &  01.02 \\
                         \thisw CSP &             Improved &  49.63 &  00.32 &  00.73 \\
             \thiswn Minimum Filter &                  MSE &  49.61 &  00.26 &  01.41 \\
             \thiswn Maximum Filter &                  MSE &  49.59 &  00.19 &  01.17 \\
                  \thiswn Down \& Up &                 SSIM &  49.51 &  00.32 &  01.12 \\
\bottomrule
\end{tabularx}

\vspace{0.15em}
	\caption{Detection performance [\%] in local scenario in terms of 
	accuracy (average + standard dev.) and false positives (average)
	over all scaling algorithms and libraries.
	$\,$ \thisws Proposed in this paper.
	}
	\label{tab:eval_limited_modification}
\end{table}

\paragraph{Analysis}
Most methods of the spatial paradigm are not effective anymore. The
reason is that scaling attacks change only a small area. Thus, 
most of the compared areas still correspond to the 
original image, making a comparison difficult.  On the contrary, the 
frequency approach is still effective, because even a limited, 
modified image area causes periodic frequency peaks.  However, the 
impact on the frequency spectrum is weaker, so that the 
performance decreases compared to attacks modifying the full image.

\paragraph{Ensemble of Detection Methods}
We also study ensembles of detection methods with local 
modifications. For the best-per-paradigm ensemble, we 
consider peak spectrum and the targeted patch-clean 
filter. 
The $K$-best ensembles consist of the first $K$ entries in
\autoref{tab:eval_limited_modification}.
We consider $K=\lbrace3, 4\rbrace$, as we have only four effective 
approaches (see \autoref{tab:eval_limited_modification}).

\autoref{tab:eval_local_modification_ensemble} shows the performance.
With majority voting and $K$-best ensembles, the false-positive rate
can be reduced to~0\% by sacrificing some accuracy.  In turn, the
combination of majority voting and best-per-paradigm slightly improves
the accuracy, but with more false positives. Note, however, that the
peak-spectrum method alone would achieve an average accuracy of 
91.76\% at a
comparable false-positive rate of 1.85\%.  Taken together, no ensemble
outperforms the individual methods in any aspect. In terms of
performance, the ensemble seems not directly beneficial. Yet, it
provides benefits against adaptive attacks as we will see in
\autoref{sec:adaptiveattack}.

\paragraph{Comparing Scaling Algorithms and Libraries}
In \appref{sec:appendix-evaluation-scalesetups}, we analyze the 
individual 
detection performance. The library has no effect, but we observe a 
duality with more advanced scaling algorithms: Frequency methods 
become better while clean-signal methods become worse. With local 
modifications, a defender should therefore choose the detection 
method based on the scaling algorithm.

\begin{table}[t]
	\centering
	\setlength\tabcolsep{3pt} %
	\renewcommand{\arraystretch}{1.2}%
	\begin{tabularx}{1\linewidth}{lRRRRRR}
		\toprule
		& \multicolumn{3}{c}{Majority} & \multicolumn{3}{c}{One 
		Winner 
			Takes All} \\
		\cmidrule(lr){2-4} \cmidrule(lr){5-7}
		Ensemble & Acc & TPR & FPR & Acc & TPR & FPR \\
		\midrule
		Best Per Paradigm & 92.00 $\pm$~1.57& 86.01 $\pm$~4.00 & 2.00 
		$\pm$~1.19 & --- & --- & --- \\	
		$K$ Best ($K$=3) & 84.84 $\pm$~1.96& 69.67 $\pm$~3.92 & 0.00 
		$\pm$~0.00 & 91.66 $\pm$~1.72& 86.49 $\pm$~3.52& 3.17 
		$\pm$~0.43\\
		$K$ Best ($K$=4) & 86.35 $\pm$~1.45& 72.70 $\pm$~2.90 & 0.00 
		$\pm$~0.00 & 91.10 $\pm$~1.96& 87.08 $\pm$~2.81& 4.88 
		$\pm$~1.38\\
		\bottomrule
	\end{tabularx}
	\vspace{0.15em}
	\caption{Ensemble in local scenario. 
		Each cell shows the average and standard deviation over all  
		scaling libraries and algorithms.}
	\label{tab:eval_local_modification_ensemble}
	\vspace{-0.5cm}
\end{table}

\paragraph{Varying Backdoors}
Next, we study the detection performance with more backdoors that 
differ in type and location.
In addition to our previously used backdoor (a black 
\emph{box} in the lower left corner), we examine (i) a black 
\emph{circle} embedded in the upper right corner, and (ii) a 
\emph{rainbow}-like box~\cite{LiuMaAaf+18} embedded in the 
lower left corner. 
These backdoors allow us to study the impact of shape, filling, and 
location. The box and rainbow patterns, for instance, are common 
patterns in backdoor attacks~\cite{LiuMaAaf+18, SalWenBac22, 
QuiRie20}. The first two columns in 
\autoref{fig:appendix_eval_backdoor_examples} in the Appendix show 
examples for all backdoor types.

\autoref{tab:eval_local_modification_ablation_samebackdoor_traintest} 
shows the performance. 
While the box and circle are well detectable, the rainbow-backdoor is 
only detected in 3 of 4 cases. We attribute this to the mixed 
filling, which causes weaker peaks in the frequency spectrum. 
Moreover, for three methods, the circle backdoor is slightly better 
detectable than the box. This is because the circle is 
larger, consuming 305~pixels compared to 225~pixels by the box. 
We verified this by reducing the circle size. The detection rates 
then become similar to the box backdoor.
In summary, we study multiple backdoor types 
with scaling attacks. Our approaches detect 
them, yet the performance relies on the backdoor~type.  

\paragraph{Varying Backdoors At Train--Test Time}
So far, the training dataset for calibrating the detection methods 
have had the same backdoor as the test dataset. In practice, this 
might not be realistic. As a remedy, we study an additional scenario 
where the training dataset is calibrated on a different backdoor than 
the one used in the test dataset. We study the same three backdoor 
types as before.

\autoref{tab:eval_local_modification_ablation_study_cross_peak_spectrum}
shows the detection rate of the peak-spectrum 
method as a matrix for all possible backdoor combinations during 
train and test time.
Using different backdoors has no impact on 
the detection. The values in each column in 
\autoref{tab:eval_local_modification_ablation_study_cross_peak_spectrum}
are almost identical, irrespective of the backdoor at training time. 
The other detection methods have the same behavior and are 
reported in 
\autoref{tab:eval_local_modification_ablation_study_cross_appendix} 
in \appref{sec:appendix-evaluation-varbackdoors}. %
We further describe the reasons for these results in 
\appref{sec:appendix-evaluation-varbackdoors}.

\begin{table}[t]
	\centering
	\begin{tabularx}{\columnwidth}{llll}
\toprule
                      Method &                Box &             Circle &            Rainbow \\
\midrule
               Peak Spectrum &  89.81 $\pm$ 04.75 &  93.03 $\pm$ 02.82 &  78.77 $\pm$ 05.85 \\
               Peak Distance &  80.98 $\pm$ 02.83 &  87.79 $\pm$ 02.22 &  68.36 $\pm$ 02.81 \\
 Targeted Patch-Clean Filter &  76.38 $\pm$ 12.35 &  70.55 $\pm$ 14.26 &  67.89 $\pm$ 22.53 \\
          Patch-Clean Filter &  75.72 $\pm$ 04.91 &  79.50 $\pm$ 05.93 &  68.50 $\pm$ 06.42 \\
\bottomrule
\end{tabularx}

\vspace{0.15em}
	\caption{Detection performance with varying backdoors (accuracy 
	$\pm$ standard deviation). Only the four
	effective detection methods are shown.}
	\label{tab:eval_local_modification_ablation_samebackdoor_traintest}
\end{table}

\begin{table}
	\centering
	\begin{tabularx}{\columnwidth}{XXXX}
		\toprule
		& \multicolumn{3}{c}{Backdoor Test-Time}
		\\
		\cmidrule(lr){2-4}	
		Backdoor \mbox{Train-Time}	& Box & Circle & Rainbow \\
		\midrule
		     Box &  89.81 $\pm$ 04.75 &  93.03 $\pm$ 02.80 &  78.77 $\pm$ 05.85 \\
  Circle &  89.66 $\pm$ 04.94 &  93.03 $\pm$ 02.82 &  78.35 $\pm$ 05.67 \\
 Rainbow &  89.76 $\pm$ 04.81 &  93.00 $\pm$ 02.78 &  78.77 $\pm$ 05.85 
		\\
		\bottomrule	
	\end{tabularx}
\vspace{0.15em}
	\caption{Performance of peak spectrum with varying 
	backdoors at train--test time (accuracy $\pm$ standard 
	dev.). The rows show the used backdoor at 
	training time, the columns the backdoor at test time.}
	\label{tab:eval_local_modification_ablation_study_cross_peak_spectrum}
\end{table}

Overall, we conclude that we can also detect scaling 
attacks if the backdoor---used to calibrate the detection---is 
different to the finally used backdoor from the adversary. Only the 
backdoor type itself, such as a circle or rainbow-like backdoor, 
affects the detection.

\paragraph{Summary}
Even in the challenging scenario where only a local image area is 
manipulated, a reliable detection is possible. However, only four 
approaches are effective: our frequency approaches based on
peak spectrum and peak distance, as well as both patch-clean 
filters. 
Finally, our results show that the defender does not need to have an 
exact knowledge of the employed backdoor.

\section{Adaptive Attacks}
\label{sec:adaptiveattack}
Finally, we study an adaptive attacker who is aware of the deployed 
detection method and adjusts the attack strategy accordingly. We 
examine the global and the local modification scenario again, but 
limit our analysis to only the successful methods from 
\autoref{sec:evaluation}. 
Note that we now have to analyze the detection rate \emph{and}
both goals of a scaling attack. Let \adaptiveattimg be 
the adaptive version of~\attimg, an attack has to 
fulfill:
(\goalA) $\scalefunc(\adaptiveattimg) \sim \tarimg$, and (\goalB) 
$\adaptiveattimg \sim \srcimg$. 
The second goal~\goalB is evaluated with the 
PSNR.
The first goal~\goalA is tested by computing the \emph{attack 
success rate} 
(ASR). In the global scenario, we define the ASR as the ratio of 
matches in the \mbox{top-5} predictions from VGG19 between 
\adaptiveattimg and \attimg. In the local scenario with backdoors, 
we define the ASR as the ratio of successful matches of the 
backdoor's target class in the \mbox{top-5} predictions.
\appref{sec:appendix-backdoor-setup} provides more information on the 
finetuning setup to measure the ASR with backdoors.

\subsection{Attacking the Frequency Paradigm}
To mislead a peak analysis, an adaptive attacker has to hide the 
periodic traces caused by a scaling attack. In the 
following, we analyze different methods to achieve this.

\paragraph{Suppressing Frequency}
We begin with a targeted attack against our peak-spectrum 
analysis. We shortly introduce the concept before presenting the 
empirical results.

\begin{figure}[b]
	\centering
	\includegraphics{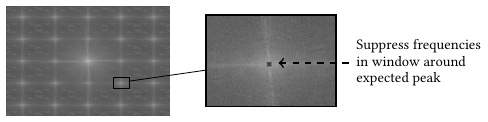}
	\vspace{-0.40em}
	\caption{Adaptive attack by suppressing frequencies with
	$\suppressfactor = 0.0$. Note %
	that such a strong reduction is often not needed.}
	\label{fig:adaptive_disable_peak_approach}
\end{figure}

\minipara{Approach}
The idea is to suppress the frequency spectrum in the 
window that is used by the defense. In particular, let \Fourier 
denote the frequency spectrum of the attack image and 
$\window \in W$ each window used by the spectrum analysis. The 
spectrum is then manipulated as follows:
\begin{align}
	\Fourier(\attimg)[\window] = \suppressfactor \cdot 
	\Fourier(\attimg)[\window] \quad \forall \; \window \in W ,
\end{align} 
where $\suppressfactor$ is a parameter to control the 
reduction and $[\cdot]$~selects a subset of frequencies. 
\autoref{fig:adaptive_disable_peak_approach} illustrates the adaptive 
attack by setting all frequencies in the window $\window$ to zero. 

\minipara{Results}
This adaptive attack has a strong impact on the detection performance, 
as \autoref{tab:adaptive_disable_peak} shows. The average detection 
accuracy notably decreases with a smaller value of $\suppressfactor$.
Note that using $\suppressfactor=0.0$ leads to a too strong 
frequency reduction, so that the detection threshold could be simply 
inverted, increasing the accuracy, for instance, to 66.63\% (100\% - 
33.37\%) in the global case.
To provide more intuition on the resulting image quality, 
\autoref{fig:appendix_adaptive_attacks_backdoor_output_disable} in 
the Appendix shows an exemplary 
image from our evaluation for varying values of 
$\suppressfactor$. 

Regarding the goals \goalA and \goalB, the adaptive attack has only 
an impact in the global scenario. Here, the ASR decreases and the 
attack images loose brightness and contrast. The output images become 
a mixture between the original and novel content. 
On the contrary, in the local scenario, 
$\suppressfactor$ has no significant impact on the ASR and the visual 
quality.
In summary, an adaptive attacker can notably decrease the detection 
performance. In the global scenario, however, the attacker has to 
sacrifice the goals \goalA and \goalB to some extent, which limits
the impact of the attack.

\begin{table}
	\centering
	\begin{tabular}{llccccc}
	\toprule
	& & \multicolumn{2}{c}{Detection}
	& \multicolumn{1}{r}{Attack \goalA} 
	& \multicolumn{2}{c}{Attack \goalB}
	\\
	\cmidrule(lr){3-4} \cmidrule(lr){5-5} \cmidrule(lr){6-7}
	Attack & Option & AvgAcc & StdAcc & ASR & AvgPSNR & 
	StdPSNR \\
	\midrule
		\multirow{6}{*}{Global} 
		         &  \suppressfactor = 1.0 & 99.85 & 00.14 & 100.00 &   23.47 &   03.27 \\
         &  \suppressfactor = 0.8 & 99.59 & 00.45 &  99.56 &   22.85 &   02.98 \\
         &  \suppressfactor = 0.6 & 98.06 & 01.49 &  97.39 &   21.00 &   03.10 \\
         &  \suppressfactor = 0.4 & 92.14 & 04.53 &  92.01 &   19.34 &   03.23 \\
         &  \suppressfactor = 0.2 & 63.10 & 08.50 &  83.71 &   18.00 &   03.25 \\
         &  \suppressfactor = 0.0 & 33.37 & 00.00 &  72.63 &   16.95 &   03.21 \\

		\cmidrule{2-7}
		\multirow{6}{*}{Local} 
		         &  \suppressfactor = 1.0 & 87.05 & 06.35 & 77.64 &   43.74 &   05.91 \\
         &  \suppressfactor = 0.8 & 82.23 & 07.56 & 77.51 &   41.25 &   07.37 \\
         &  \suppressfactor = 0.6 & 75.06 & 08.33 & 77.34 &   38.61 &   08.19 \\
         &  \suppressfactor = 0.4 & 64.56 & 08.46 & 77.44 &   36.34 &   08.63 \\
         &  \suppressfactor = 0.2 & 48.50 & 06.49 & 77.42 &   34.41 &   08.89 \\
         &  \suppressfactor = 0.0 & 33.06 & 00.32 & 77.50 &   32.75 &   08.99 \\

		\bottomrule
	\end{tabular}
\vspace{0.15em}
	\caption{Adaptive attack against peak-spectrum approach by 
		suppressing frequency peaks (Acc.\ and ASR in [\%], and PSNR 
		in [dB]).}
	\label{tab:adaptive_disable_peak}
\end{table}

\paragraph{Add Frequency Peak}
We continue with an attack against the peak-distance analysis. 
Again, we shortly introduce the concept before presenting results.

\begin{figure}
	\centering
	\includegraphics{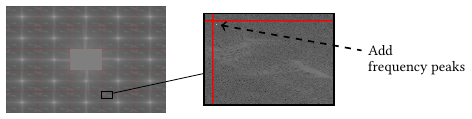}
	\vspace{-0.40em}
	\caption{Adaptive attack by adding peaks with
		$\addpeakfactor = 25$. Note that our evaluation shows 
		that such a strong addition is often not necessary.}
	\label{fig:adaptive_add_peak_approach}
\end{figure}

\minipara{Approach}
The idea is to insert an additional peak in each excerpt so that the 
distance between the expected and maximum peak increases. Let $p \in 
P$ be the frequency location for adding a peak in each 
excerpt, the spectrum is modified as follows:
\begin{align}
	\Fourier(\attimg)[p] = \addpeakfactor^{-1} \cdot 
	\max{(\Fourier(\attimg))} \quad \forall \; p \in P ,
\end{align}
where \addpeakfactor controls the strength of the added peak. For 
simplicity, the peak locations $p \in P$ are the corners of each 
excerpt. 
\autoref{fig:adaptive_add_peak_approach} illustrates the adaptive 
attack. This frequency modification is hard to perceive in 
the whole frequency spectrum, as only very nuanced changes are 
performed.

\minipara{Results}
\autoref{tab:adaptive_add_peak} shows the results of the adaptive 
attack. The peak addition has a strong impact on the detection 
performance.
Already $\addpeakfactor=75$ notably decreases the performance in the 
global and the local modification scenario. 
At the same time, the peak addition has only a minor effect on the 
goal \goalA. To better explain the effect on~\goalB, 
\autoref{fig:appendix_adaptive_attacks_backdoor_output_add} in 
the Appendix shows an example from our evaluation. 
As $\addpeakfactor$ decreases and the peaks become stronger, the 
images loose brightness and contrast. Still, the image content 
remains intact---even at a high modification factor with 
$\addpeakfactor=25$. 
We thus conclude that \goalB is also partly achieved.
All in all, we identify a suitable range of $50 
\leqslant \addpeakfactor \leqslant 75$ where the peak-distance 
approach does not detect an attack and both goals of a scaling-attack 
are satisfied.

\begin{table}
	\centering
	\begin{tabular}{llccccc}
	\toprule
	& & \multicolumn{2}{c}{Detection}
	& \multicolumn{1}{r}{Attack \goalA} 
	& \multicolumn{2}{c}{Attack \goalB}
	\\
	\cmidrule(lr){3-4} \cmidrule(lr){5-5} \cmidrule(lr){6-7}
	Attack & Option & AvgAcc & StdAcc & ASR & AvgPSNR & 
	StdPSNR \\
	\midrule
		\multirow{3}{*}{Global} 
		         &  \addpeakfactor = 75 & 61.88 & 02.92 & 99.32 &   19.39 &   02.62 \\
         &  \addpeakfactor = 50 & 50.29 & 03.29 & 97.70 &   18.23 &   02.73 \\
         &  \addpeakfactor = 25 & 37.59 & 02.00 & 90.11 &   16.22 &   02.72 \\

		\cmidrule{2-7}
		\multirow{3}{*}{Local} 
		         &  \addpeakfactor = 75 & 33.03 & 00.26 & 82.73 &   22.83 &   05.00 \\
         &  \addpeakfactor = 50 & 33.03 & 00.26 & 84.49 &   20.54 &   04.54 \\
         &  \addpeakfactor = 25 & 32.98 & 00.31 & 88.64 &   17.25 &   03.67 \\

		\bottomrule
	\end{tabular}
	\vspace{0.15em}
	\caption{Adaptive attack against the peak-distance approach by 
		adding frequency peaks (Acc.\ and ASR in [\%], and PSNR in 
		[dB]).}
	\label{tab:adaptive_add_peak}
\end{table}

\paragraph{JPEG Compression}
As a baseline, we consider a generic and simple counterattack by 
examining JPEG compression. It is known to have an impact on 
resampling detectors in multimedia forensics~\cite{KirBoe08}, which 
analyze the frequency spectrum for peaks similar to our approach. 
\appref{sec:appendix-evaluation-adaptive} presents the results. 
Compression only affects the local scenario. However, the effect on 
the detection performance is smaller compared to our previous two
targeted adaptive attacks.

\subsection{Attacking the Spatial Paradigm}
Lastly, we discuss the attack surface of the other paradigm. 
For strengthening approaches based on down- and upscaling, an 
adaptive attack should not be possible. %
Upscaling algorithms do not suffer from the root cause that enables 
scaling attacks in the downscaling case. Upscaling algorithms usually 
use each pixel multiple times to compute the larger image, so that an 
attacker cannot hide a new signal. We 
verified the implementation of the imaging libraries OpenCV, Pillow, 
and tf.image (TensorFlow) and observed this behavior in their 
upscaling algorithms. 
Thus, if the downscaling reveals another content, the 
upscaled version will necessarily keep this content, making an 
adaptive attack difficult.

The clean-signal driven approaches based on prevention filters inherit
the security properties of the respective filter. Prior work has 
demonstrated that these filters withstand adaptive 
attackers~\cite{QuiKleArp20}. In the global scenario, we can 
thus conclude that the detection is robust. In 
the local scenario, however, the cleaning 
approach only works with a patch extraction. In this case, this patch 
extraction can introduce a new vulnerability. For example, an 
attacker could distribute a backdoor over multiple patches. This 
requires designing new backdoor methods which is beyond the scope of 
this paper. %

\subsection{Summary}
Our analysis shows that the frequency paradigm, although the strongest
under a static attacker, can be circumvented by an adaptive
attacker. On the contrary, the spatial paradigm withstands adaptive
attacks by design. Hence, we are faced with a trade-off
where the frequency and spatial paradigms complement each other in 
detection capabilities and robustness, respectively.

\section{Related Work}
\label{sec:relatedwork}
Image-scaling attacks are a novel threat to the security of
machine-learning systems. As a result, there exists only a small body 
of related work that is discussed in the following.

\minipara{Attacks}
\citet{XiaCheShe+19} have initially introduced image-scaling attacks.
\mbox{\citet{QuiKleArp20}} perform an in-depth analysis of scaling 
attacks and identify their root cause. We build our defenses on this 
understanding. 
\citet{CheSheWan+20} extend the original attack by studying 
different norms for \autoref{eq:opti_problem_basic}. Yet, these norms 
do not affect the attack's working principle and thus our proposed 
defenses.
\citet{QuiRie20} examine the application for the poisoning 
and backdoor scenario. 
This work motivates our inclusion of the 
local-modification scenario. %
Finally, \citet{GaoShuFaw22} combine adversarial examples and 
scaling attacks. We excluded the attack, since it operates in a 
different threat scenario. The attack is dependent on the learning 
model and requires an iterative adversarial-example process. On the 
contrary, we focus on general scaling attacks that are 
model-agnostic and just create the target as scaling output. 

\minipara{Defenses} To fend off scaling attacks, 
we can either \emph{prevent} or \emph{detect} an attack.
In the former case, \citet{QuiKleArp20} have extensively 
studied prevention defenses.
In the latter case, \citet{XiaCheShe+19} and \citet{KimAbuGao+21} 
have presented first ideas, evaluated with non-adaptive attackers. We 
include these approaches in our comparison, but our evaluation shows 
that they are often ineffective.

\minipara{Adversarial Learning} 
Scaling attacks are \emph{preprocessing attacks}~\cite{QuiKleArp20} 
that represent a new type of ML attack in 
addition to existing attacks, such as adversarial examples and 
poisoning~\cite{BigRol18, PapMcSin+18, QuiMaiRie19}.

\section{Conclusion}
\label{sec:conclusion}

This paper is the first comprehensive study on the detection of
image-scaling attacks. We examine the problem from multiple viewpoints
by considering various scaling algorithms, levels of modification, and
attacker models. We systematize the detection  and derive novel 
approaches based on our improved
understanding.

The frequency paradigm is the strongest detection approach in any
image-modification scenario under a static attack. Under an adaptive
attack, it is not robust and vulnerable to evasion.  The spatial
paradigm is suitable for spotting global manipulations but lacks
accuracy for local changes. However, it enables a robust detection
under adaptive attacks. Therefore, our results motivate that both
paradigms should be used as ensemble to complement each other.

Finally, we emphasize that detection should not be seen as replacement
for prevention defenses. In fact, both concepts best operate in
combination. While prevention methods block attacks during training
and inference, approaches for detection help identify on-going attacks
and ultimately help tracking down adversaries.  As a result, by
combining both concepts, we can improve the security of
machine-learning systems.

\begin{acks}
The authors would like to thank Rainer Böhme, Pascal Schöttle, and 
Daniel Arp for the discussions about the detection of scaling 
attacks. 
This work was funded by the Deutsche Forschungsgemeinschaft (DFG, 
German Research Foundation) under Germany's Excellence Strategy -- 
EXC 2092 CASA -- 390781972), the German Federal Ministry of Education 
and Research under the grant BIFOLD23B, and
the European Research Council (ERC) under the consolidator grant 
MALFOY (101043410).
Moreover, this work was supported by a fellowship within the IFI 
program of the German Academic Exchange Service (DAAD) funded by the 
Federal Ministry of Education and Research (BMBF).
\end{acks}

{	
	\footnotesize 
	\bibliographystyle{abbrvnat}

}

\appendix
\section{Frequency Analysis}
\label{sec:appendix-frequency-analysis}

In the following, we derive the computation of the peak positions 
that an image-scaling attack inevitably introduces. 
We start with an uncentered view of the frequency spectrum as well as 
two assumptions that will be gradually relaxed: the use 
of nearest scaling and an integer as scaling ratio. 
This enables us to define a simplified image model.
It mimics the blocking artifacts that a scaling attack will add in a 
periodic interval %
(recall the root-cause analysis in 
\autoref{subsec:background-scaling-attacks}). 
In particular, we denote by \mbox{$B \in \lbrace 0, 1 \rbrace ^{m 
\times n}$} an image where all values are set to 1, except for those 
on a grid in the interval~$\Tm \in \mathbb{N}$ along the vertical 
direction and in the interval~$\Tn \in \mathbb{N}$ along the 
horizontal direction:
\begin{align}
	B(i,j) =
	\begin{cases}
		0 & \text{if } i = l_1  \Tm \land j = l_2  \Tn \\
		1 & \text{otherwise},
	\end{cases}
	\\ \text{with} \;
	\; 0 \leqslant l_1 < m', \; 0 
	\leqslant l_2 < n', \; l_1, \; l_2 \in \mathbb{N} . 
	\nonumber
\end{align}

The function $B$ corresponds to the image model for JPEG 
compression by \citet{CheHsu11}, except for a different periodicity 
between the peaks in the grid. Substituting the JPEG periodicity of~8 
by the periodicity \Tm and \Tn,
we expect to observe peaks around 
each \mbox{$k_1 m'$-th} and $k_2 n'$-th 
position of the frequency spectrum if an image is manipulated by an 
image-scaling attack.
More formally, we can define the 
following binary function \mbox{$\Gamma \in \lbrace 0, 1 \rbrace ^{m 
		\times n}$} that shows at which frequency coefficient a peak 
		occurs:
\begin{align}
	\Gamma(u,v) =
	\begin{cases}
		1 &
		(u, v) = (k_1 m', k_2 n') \\
		0 & \text{otherwise}.
	\end{cases}
	\label{eq:peak-function-appendix-uncentered}
	\\
	\text{with} 
	\; 0 \leqslant k_1 < \Tm, \; 0 \leqslant k_2 < 
	\Tn, \; k_1, k_2 
	\in \mathbb{N}. \nonumber
\end{align}
To bring this into a centered view, we shift the coordinates and 
get:
\begin{align}
	\Gamma(u,v) =
	\begin{cases}
		1 &
		(u, v) = (c_m + k_1 \cdot m', c_n + k_2 \cdot n') \\
		0 & \text{otherwise}.
	\end{cases}
	\label{eq:peak-function-appendix-centered}
	\\
	\text{with} 
	\; 
	-\tfrac{\Tm}{2} \leqslant k_1 < \tfrac{\Tm}{2}, \; 
	-\tfrac{\Tn}{2} \leqslant k_2 < \tfrac{\Tn}{2}, \; 
	k_1, k_2 \in \mathbb{N}. \nonumber
\end{align}
The constants $c_m$ and $c_n$ are the index of the spectrum's middle.

Let us now relax the assumptions. First, although scaling 
attacks manipulate more pixels for other algorithms such as bilinear 
and bicubic scaling, their manipulation still operates on a grid. 
Hence, \autoref{eq:peak-function-appendix-centered} can also be 
applied in these cases.
Next, we relax the integer assumption of the scaling ratio. In 
practice, the ratio can also be a rational number. 
A closer analysis shows that the step width~\Tm and \Tn can alternate 
in this case. As a result, we observe an additional periodic signal. 
The frequency spectrum has additional sub-peaks. Still, 
\autoref{eq:peak-function-appendix-centered} is also applicable in 
this case, as the major step widths correspond to \Tm and \Tn. Yet, 
we set 
$k_1 \in \mathbb{N}$ and $k_2 \in \mathbb{N}$ in 
\autoref{eq:peak-function-appendix-centered} as follows:  
\mbox{$
-\tfrac{\Tm}{2} \leqslant k_1 \leqslant \tfrac{\Tm}{2}, 
\; 
-\tfrac{\Tn}{2} \leqslant k_2 \leqslant \tfrac{\Tn}{2} 
$}.

\section{Frequency Peaks in Natural Images}
\label{sec:appendix-peaks-benign-images}

\begin{figure}[h]
	\centering
	\includegraphics{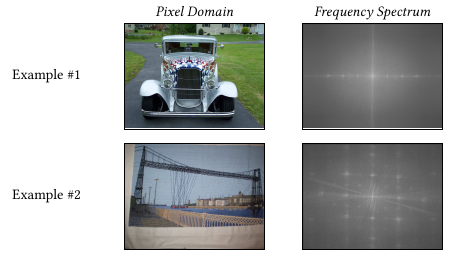}
	\vspace{-1em}
	\caption{Unmodified examples from ImageNet with natural peaks.}
	\label{fig:peaks_benign_appendix}
\end{figure}

In \autoref{fig:peaks_benign_appendix}, we show two unmodified, 
benign image examples from ImageNet 
where the frequency domain naturally contains frequency 
peaks. The CSP approach would flag both images as attack. In the 
first image, for example, the periodic pattern of the radiator grill 
causes the peaks. 

\section{Patch-Clean Filter}
\label{sec:appendix-patch-local-detection}
Here, we provide more details on the \emph{patch-clean filter}.
Recall that we create two versions of \attimg. First, 
we downscale \attimg directly using the vulnerable 
scaling algorithm, yielding \outimg. Second, we apply a prevention 
filter $\filter$ on \attimg and downscale $\attimg' = 
\filter(\attimg)$, yielding $\outimg'$. 
Hence, the adversarial modifications are present in \outimg, while 
$\outimg'$ is a clean~version. 
We only use the median filter for $\filter$ in our evaluation. 
The filter better preserves the visual quality, which is important 
when small patches are used for comparison.

We apply a Gaussian filter on \outimg to smooth it slightly, as
$\outimg'$ is slightly smoothed due to the previous application of 
$\filter$. For simplicity, we fix the Gaussian 
kernel size to (3,3) and the kernel standard deviation to 0. 
Finally, we divide both images into patches.  
Let $\outimg_{i}$ and~$\outimg'_{i}$ denote corresponding
patches from the same region in \outimg and~$\outimg'$.  Let $L$ be 
the number of patches. We compute
$v_i = \mathrm{PSNR}(\outimg_{i}, \outimg'_{i}) \; \forall i = 1,
\dots, L$.  The final detection score is given as:
\begin{align}
	\vert \mathrm{mean}(\lbrace v_i \rbrace) - \mathrm{min}(\lbrace 
	v_i \rbrace) \vert.
\end{align}  
If a scaling attack changes only a local area, only a few patches 
will have an unusually small PSNR 
value. Compared to benign images, the minimum over all patches is 
then significantly smaller than the mean. The
difference between mean and minimum thus reveals local attacks.
To obtain patches, we iterate over the image with a sliding 
window and extract sub-windows as patches. The sub-window size~$\w$ 
in each direction (total side length $2 \cdot \w$)
and the stride~$\stride$ are two parameters that we 
determine on a validation set in our evaluation.

Note that we have tested
more advanced approaches to extract patches, such as k-means based
segmentation or the selective search image segmentation algorithm
proposed by \citet{ChoTraPel20} in the backdoor context. While the
former often misses the backdoor region, the latter creates a too 
exhaustive list of possible regions that increases the number of 
false positives.

\section{Backdoor Evaluation Setup}
\label{sec:appendix-backdoor-setup}
Here, we describe the different setups to check that our backdoors 
are effective. 

\paragraph{Static Adversary}
We evaluate two settings to ensure that the backdoors work with 
regular image-scaling attacks.

\vspace{0.3em}
\minipara{Training from Scratch}
As testing a backdoor attack by training VGG19 from scratch
is computationally expensive, 
we resort to an equivalent, but simpler setup with the 
\mbox{CIFAR-10} dataset~\citep{KriHin09} and the neural network 
architecture from \citet{CarWag17}. 
In particular, we use 40,000 CIFAR images for training and embed a 
box backdoor on a varying number of training samples. We use the 
CIFAR test set for evaluating the clean accuracy on unmodified 
instances, and for measuring the attack success rate after embedding 
a backdoor. The latter shows how often a backdoored image triggers 
its target class. Modifying 1\% of the training samples leads to a 
success rate of 68\% while 10\% lead to a success rate of 97\%. 
The clean accuracy is not largely affected.
Next, we evaluate the attack success rate when applying scaling 
attacks to hide the backdoor on test samples.
The difference with and without scaling attack is less than 1\%. 
We conclude that our backdoors are effective. Scaling attacks have no 
considerable impact on the backdoor effectivity.

\vspace{0.3em}
\minipara{Finetuning}
To check the validity on VGG19 directly, we additionally test a 
finetuning setup to 
embed a backdoor. To this end, our training set consists of 
350~images from the 585 backdoored images where a scaling attack is 
used to hide the backdoor (see~\autoref{subsec:evaluation-setup}).
In addition, we collect 2,000 novel, unmodified images from ImageNet.
For the backdoor, we choose a random target class as label. 
For finetuning, we use Adam with a small 
learning rate of 1e-6. 
To check the performance, we collect 2,000 further ImageNet images as 
benign test set and the remaining 235~backdoored images as attack 
test set.
The former set is used to get the top-5 accuracy before and after 
finetuning. The latter set is used to measure the attack success 
rate, that is, how often a backdoored image can activate the target 
class in the \mbox{top-5} predictions.
We report the average and standard deviation over 10 random target 
classes and over all scaling libraries \& algorithms.

The attack success rate is 75.66\% $\pm$ 6.96\%, underlining that a 
backdoor in combination with scaling attacks can be effectively 
embedded with a very simple finetuning setup. 
The accuracy on the benign test set drops from 95.70\% to 
93.48\% $\pm$ 0.39\%.
Overall, we can conclude that our backdoor setup---with scaling 
attacks to hide backdoors---is effective. 

\paragraph{Adaptive Adversary}
To measure the success rate of backdoors after applying the
adaptive attacks from \autoref{sec:adaptiveattack}, we adopt the 
prior finetuning setup. Yet, we now use the adaptive version of each 
attack image that 
contains the backdoor.
For each adaptive attack with respective 
parameters, we run a separate finetuning setup to embed the 
backdoor.

We report the attack success rate for each setup in 
\autoref{tab:adaptive_disable_peak} and 
\autoref{tab:adaptive_add_peak} in \autoref{sec:adaptiveattack}.
The accuracy drop on the benign test set is comparable to the static 
attack before. The accuracy after finetuning over all adaptive 
attacks is 93.40\% $\pm$ 0.14\%. Due to this very low deviation 
across 
all adaptive attacks, we omit the clean accuracy in the tables in 
\autoref{sec:adaptiveattack}.

\section{Hyperparameters}
\label{sec:appendix-hyperparams}
Peak spectrum uses $\w=5$, the patch-clean filter uses 
$(\w=22, \stride=11)$, and the targeted patch-clean filter uses 
$(\w=11, \stride=11, q=0.6)$.

\section{Comparing Scaling Setups}
\label{sec:appendix-evaluation-scalesetups}
We have presented aggregated results over all scaling 
algorithms and libraries so far. Yet, scaling attacks have to modify 
more pixels for scaling algorithms with larger kernels such as 
bilinear or bicubic scaling. In this section, we therefore analyze 
whether specific algorithms and libraries have an impact on the 
detection performance. 

\paragraph{Global Modification}
\autoref{tab:eval_full_image_compare_algs_libs_tf} shows the 
detection performance for TensorFlow per scaling algorithm. 
The OpenCV results are similar and omitted due to lack of space. 
Our proposed frequency methods work for all algorithms equally well.
The performance of some clean-signal driven approaches 
decrease with bilinear and bicubic scaling.
We attribute this to the increased number of necessary pixel 
reconstructions by the filter. This affects the visual quality 
and thus the image comparison. 
The PSNR and the random filter are then more affected than 
the SSIM and the median filter. 

\paragraph{Local Modification}
\autoref{tab:eval_local_image_compare_algs_libs_tf} shows the 
results for TensorFlow. 
The results for OpenCV are similar and shown in
\autoref{tab:eval_local_image_compare_algs_libs_cv}.
We observe a duality with more advanced algorithms: The frequency
methods become better while the clean-signal methods
become worse. With larger kernels, an attack has to modify more
pixels. This is advantageous for frequency methods where the peaks
become more prevalent. In contrast, more pixels have to be
reconstructed with the filter-based methods, making a comparison
more difficult.

\section{Backdoors and Frequency Spectrum}
\label{sec:appendix-evaluation-varbackdoors}
\autoref{tab:eval_local_modification_ablation_study_cross_appendix} 
shows the detection accuracy as a matrix for all 
backdoor combinations during training and test time. Peak spectrum 
is already shown in the main section in 
\autoref{tab:eval_local_modification_ablation_study_cross_peak_spectrum}.
Different backdoors at train-test time do not affect the performance.
For the frequency analysis, the reason is that the peak positions do 
not depend on the backdoor's location or shape in the pixel domain. 
The frequency peaks depend on the distance between the periodic 
changes. 
Neither are the patch-based approaches affected. The compared patches 
are only derived from the current input image, so that 
a backdoor will cause an observable difference in a patch. 
Thus, varying locations and shapes of backdoors at test time are here
detectable, too.

\begin{figure}[h]
	\centering
	\includegraphics{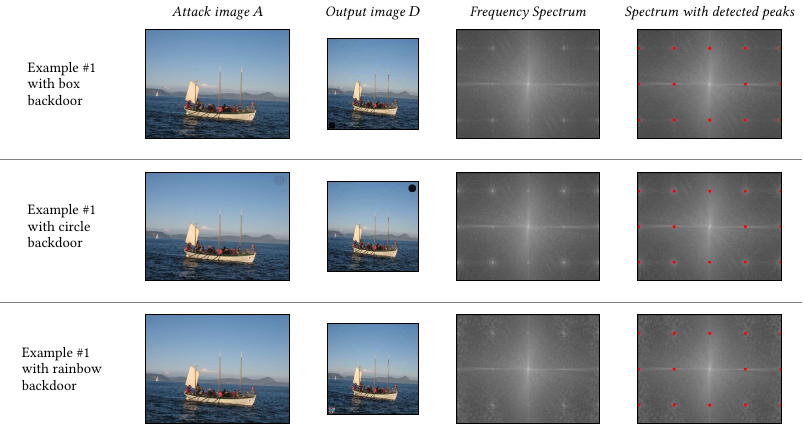}
	\caption{Evaluation examples from backdoor detection. The columns 
	show an attack image, its downscaled version, the frequency 
	spectrum of \attimg, and the detected peaks with our frequency 
	method. The first row depicts a scaling attack with a box 
	backdoor in the bottom-left corner. The row in the middle 
	shows an attack with a circle backdoor in the upper-right corner. 
	The last row depicts an attack with the rainbow backdoor in 
	the bottom-left corner. The plots highlight that the frequency 
	traces do not depend on the backdoor's shape or location.}
	\label{fig:appendix_eval_backdoor_examples}
\end{figure}

\autoref{fig:appendix_eval_backdoor_examples} shows examples from our 
evaluation if different backdoors are embedded in combination with a 
scaling attack. The figure also shows the respective frequency 
spectrum. 

\section{Adaptive Attacks}
\label{sec:appendix-evaluation-adaptive}
In this section, we present additional results for the adaptive 
attacks.

\autoref{tab:adaptive_jpeg_spectrum_appendix} and
\autoref{tab:adaptive_jpeg_distance_appendix} show the results of 
JPEG compression as adaptive attack.
The global scenario is robust, while the local scenario is more 
affected. As strong JPEG compression removes high frequencies, it 
affects the local case with weaker scaling-attack peaks more.
Still, the peak-spectrum method can detect more than 70\% even with 
strong compression. Compared to our targeted attacks in 
\autoref{sec:adaptiveattack}, compression is less effective.

\begin{figure*}
	\centering
	\includegraphics{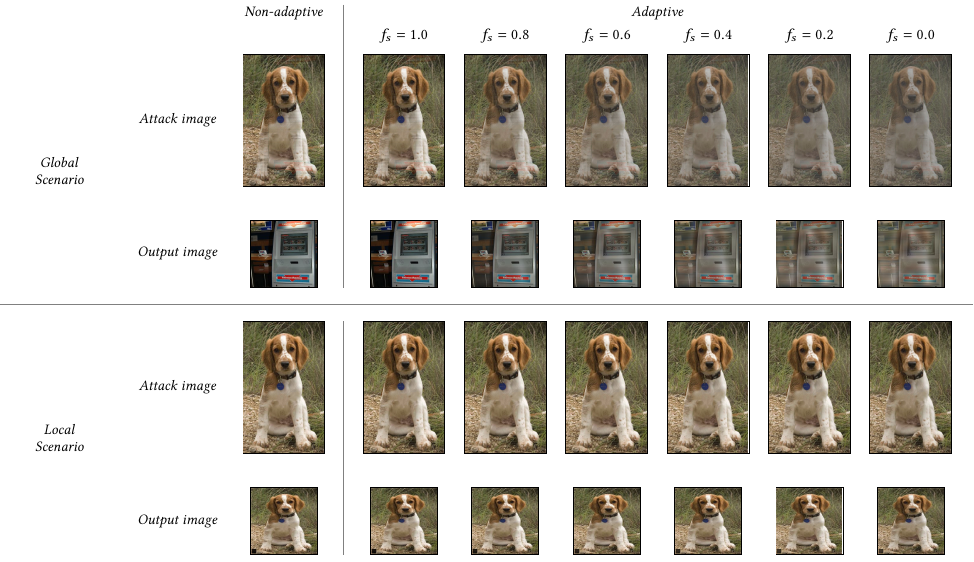}
	\caption{Evaluation examples for the adaptive attack based on 
		suppressing peaks. 
		The first column shows the non-adaptive, 
		original scaling attack, while the further columns show the 
		adaptive modification to bypass the detection. 
		It is visible that small values of 
		\suppressfactor have a clear impact in the global scenario.
	}
	\label{fig:appendix_adaptive_attacks_backdoor_output_disable}
\end{figure*}

\begin{figure*}
	\centering
	\includegraphics{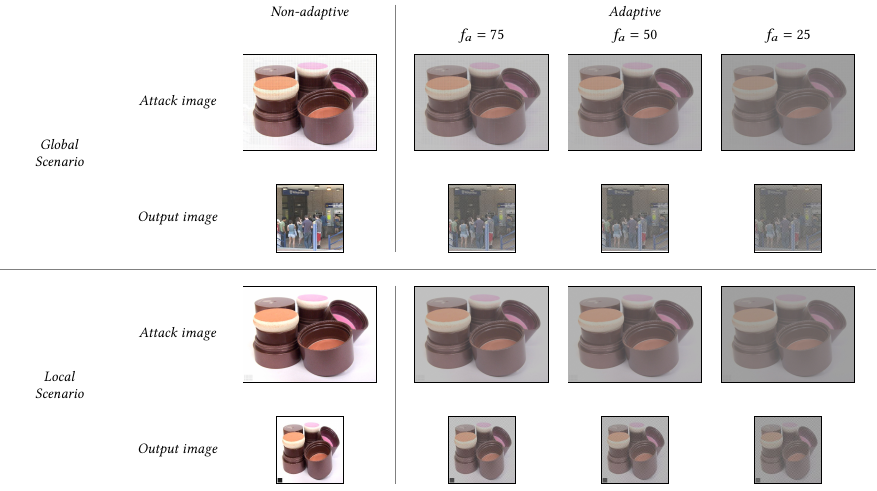}
	\caption{Evaluation examples for the adaptive attack based on adding 
		peaks. 
		The first column shows the non-adaptive, 
		original scaling attack, while the further columns show the 
		adaptive modification to bypass the detection. 
		It is visible that adding peaks reduces the brightness, but 
		the image content remains intact.
	}
	\label{fig:appendix_adaptive_attacks_backdoor_output_add}
\end{figure*}

\autoref{fig:appendix_adaptive_attacks_backdoor_output_disable} and 
\autoref{fig:appendix_adaptive_attacks_backdoor_output_add} show 
examples from the evaluation to provide more insights on the visual 
quality.

\begin{table}
	\centering
	\begin{tabularx}{\columnwidth}{llRRR}
\toprule
                     \thiswn Method &               Option &  Nearest &  Linear &   Cubic \\
\midrule
               \thisw Peak Distance &                      &  100.00 & 100.00 & 100.00 \\
               \thisw Peak Spectrum &                      &  100.00 & 100.00 & 100.00 \\
                \thisw Clean Filter &  Median filter, SSIM &  100.00 &  99.83 &  99.66 \\
                  \thiswn Down \& Up &            Histogram &  100.00 &  97.78 &  98.29 \\
                   \thisw Down \& Up &                 PSNR &   97.44 & 100.00 &  96.25 \\
                  \thiswn Down \& Up &                  MSE &   97.44 & 100.00 &  96.25 \\
                \thisw Clean Filter &  Random filter, SSIM &   99.66 &  99.49 &  96.42 \\
 \thisw Targeted Patch-Clean Filter &                      &   99.15 &  98.81 &  93.69 \\
                \thisw Clean Filter &  Median filter, PSNR &   98.81 &  97.78 &  87.20 \\
             \thiswn Maximum Filter &                 SSIM &   85.15 &  87.37 &  91.98 \\
             \thiswn Minimum Filter &                 SSIM &   84.47 &  85.67 &  89.59 \\
                \thisw Clean Filter &  Random filter, PSNR &   88.05 &  86.52 &  71.33 \\
                            \midrule \thiswn Average &                      &   95.85 &  96.10 &  93.39 \\
\bottomrule
\end{tabularx}

	\vspace{0.15em}
	\caption{Detection accuracy per scaling algorithm in 
		TensorFlow in the global-modification scenario. 
		Only approaches with AvgAcc > 80\% in 
		\autoref{tab:eval_full_image_modification}
		are presented.
		}
	\label{tab:eval_full_image_compare_algs_libs_tf}
\end{table}

\begin{table}
	\centering
	\begin{tabularx}{\columnwidth}{llRRR}
\toprule
                     \thiswn Method & Option &  Nearest &  Linear &  Cubic \\
\midrule
               \thisw Peak Spectrum &        &   84.81 &  94.37 & 93.52 \\
               \thisw Peak Distance &        &   77.99 &  83.96 & 84.98 \\
 \thisw Targeted Patch-Clean Filter &        &   86.18 &  80.72 & 61.95 \\
          \thisw Patch-Clean Filter &        &   80.20 &  74.57 & 66.55 \\
                            \midrule \thiswn Average &        &   82.30 &  83.40 & 76.75 \\
\bottomrule
\end{tabularx}

	\vspace{0.15em}
	\caption{Detection accuracy per scaling algorithm in TensorFlow 
	in the local-modification scenario.
		Only the four effective approaches from 
		\autoref{tab:eval_limited_modification} are presented.
		}
	\label{tab:eval_local_image_compare_algs_libs_tf}
\end{table}

\begin{table}
	\centering
	\begin{tabularx}{\columnwidth}{llRRR}
\toprule
                     \thiswn Method & Option &  Nearest &  Linear &  Cubic \\
\midrule
               \thisw Peak Spectrum &        &   84.81 &  93.00 & 93.52 \\
               \thisw Peak Distance &        &   77.99 &  81.57 & 81.57 \\
 \thisw Targeted Patch-Clean Filter &        &   86.18 &  76.79 & 56.48 \\
          \thisw Patch-Clean Filter &        &   80.20 &  74.57 & 74.23 \\
                            \midrule \thiswn Average &        &   82.30 &  81.48 & 76.45 \\
\bottomrule
\end{tabularx}

	\caption{Detection accuracy per scaling algorithm in 
		OpenCV in the local-modification scenario. 
		Only the four effective approaches from 
		\autoref{tab:eval_limited_modification} are presented.
		}
	\label{tab:eval_local_image_compare_algs_libs_cv}
\end{table}

\begin{table}
	\centering
	\begin{tabularx}{\columnwidth}{lXXXX}
		\toprule
		& & \multicolumn{3}{c}{Backdoor Test-Time}
		\\
		\cmidrule(lr){3-5}	
	 Method & Backdoor Train-Time	& Box & Circle & Rainbow \\
		\midrule
		\multirow{3}{*}{PD} 
	 &      Box &  80.98 $\pm$ 02.83 &  87.91 $\pm$ 02.26 &  68.28 $\pm$ 02.73 \\
 &  Circle &  80.94 $\pm$ 02.62 &  87.79 $\pm$ 02.22 &  68.19 $\pm$ 02.65 \\
 & Rainbow &  81.01 $\pm$ 02.84 &  87.91 $\pm$ 02.26 &  68.36 $\pm$ 02.81 
		\\
		\cmidrule{2-5}
		\multirow{3}{*}{TPF} 
	 &      Box &  76.38 $\pm$ 12.35 &  70.70 $\pm$ 14.39 &  67.84 $\pm$ 22.35 \\
 &  Circle &  76.13 $\pm$ 12.20 &  70.55 $\pm$ 14.26 &  66.92 $\pm$ 21.28 \\
 & Rainbow &  76.11 $\pm$ 12.49 &  70.65 $\pm$ 14.47 &  67.89 $\pm$ 22.53 
		\\
		\cmidrule{2-5}
		\multirow{3}{*}{PF} 
	 &      Box &  75.72 $\pm$ 04.91 &  79.57 $\pm$ 05.92 &  68.70 $\pm$ 06.27 \\
 &  Circle &  75.62 $\pm$ 04.81 &  79.50 $\pm$ 05.93 &  68.53 $\pm$ 06.35 \\
 & Rainbow &  75.62 $\pm$ 04.96 &  79.55 $\pm$ 05.98 &  68.50 $\pm$ 06.42 
		\\
		\bottomrule
	\end{tabularx}
	\caption{Detection performance with varying 
		backdoors at train--test time (accuracy $\pm$ standard 
		deviation). The rows show the used backdoor at 
		training time, the columns the backdoor at test time. 
		Abbreviations for detection methods in first 
	column: 
	PD=Peak Distance, TPF=Targeted Patch-Clean 
	Filter, PF=Patch-Clean Filter.}
	\label{tab:eval_local_modification_ablation_study_cross_appendix}
\end{table}

\onecolumn

\begin{table}
\begin{minipage}{0.485\columnwidth}

\centering
\begin{tabular}{llccccc}
	\toprule
	& & \multicolumn{2}{c}{Detection}
	& \multicolumn{1}{r}{Attack \goalA} 
	& \multicolumn{2}{c}{Attack \goalB}
	\\
	\cmidrule(lr){3-4} \cmidrule(lr){5-5} \cmidrule(lr){6-7}
	Attack & Option & AvgAcc & StdAcc & ASR & AvgPSNR & 
	StdPSNR \\
	\midrule
	\multirow{5}{*}{Global} 
	         &          Q = 90 & 99.82 & 00.17 & 84.64 &   23.79 &   03.24 \\
         &          Q = 80 & 99.77 & 00.30 & 66.89 &   23.68 &   03.25 \\
         &          Q = 70 & 99.37 & 00.58 & 47.20 &   23.77 &   03.41 \\
         &          Q = 60 & 98.06 & 01.26 & 33.92 &   23.94 &   03.56 \\
         &          Q = 50 & 96.13 & 01.61 & 24.32 &   24.13 &   03.67 \\

	\cmidrule{2-7}
	\multirow{6}{*}{Local} 
	         &          Q = 90 & 84.82 & 05.88 & 76.98 &   41.58 &   04.43 \\
         &          Q = 80 & 81.00 & 06.03 & 76.23 &   39.51 &   04.57 \\
         &          Q = 70 & 77.40 & 06.85 & 75.31 &   37.76 &   04.23 \\
         &          Q = 60 & 74.24 & 07.23 & 74.42 &   36.37 &   03.76 \\
         &          Q = 50 & 70.92 & 07.28 & 74.00 &   35.50 &   03.84 \\

	\bottomrule
\end{tabular}
\caption{Adaptive attack against the peak-spectrum approach based 
	on compression
	(Acc.\ and ASR in [\%], PSNR in [dB], and $Q$ is 
	the JPEG compression level).}
\label{tab:adaptive_jpeg_spectrum_appendix}

\end{minipage}
\hfill
\begin{minipage}{0.485\columnwidth}

\centering
\begin{tabular}{llccccc}
	\toprule
	& & \multicolumn{2}{c}{Detection}
	& \multicolumn{1}{r}{Attack \goalA} 
	& \multicolumn{2}{c}{Attack \goalB}
	\\
	\cmidrule(lr){3-4} \cmidrule(lr){5-5} \cmidrule(lr){6-7}
	Attack & Option & AvgAcc & StdAcc & ASR & AvgPSNR & 
	StdPSNR \\
	\midrule
	\multirow{5}{*}{Global} 
	         &          Q = 90 & 99.95 & 00.06 & 84.64 &   23.79 &   03.24 \\
         &          Q = 80 & 99.95 & 00.06 & 66.89 &   23.68 &   03.25 \\
         &          Q = 70 & 99.95 & 00.06 & 47.20 &   23.77 &   03.41 \\
         &          Q = 60 & 99.82 & 00.11 & 33.92 &   23.94 &   03.56 \\
         &          Q = 50 & 99.24 & 00.44 & 24.32 &   24.13 &   03.67 \\

	\cmidrule{2-7}
	\multirow{6}{*}{Local} 
	         &          Q = 90 & 72.19 & 02.88 & 76.78 &   41.58 &   04.43 \\
         &          Q = 80 & 67.85 & 02.34 & 76.11 &   39.51 &   04.57 \\
         &          Q = 70 & 64.07 & 02.46 & 75.58 &   37.76 &   04.23 \\
         &          Q = 60 & 60.61 & 02.71 & 74.41 &   36.37 &   03.76 \\
         &          Q = 50 & 58.10 & 03.00 & 73.66 &   35.50 &   03.84 \\

	\bottomrule
\end{tabular}
\vspace{0.15em}
\caption{Adaptive attack against the peak-distance approach based 
	on compression
	(Acc.\ and ASR in [\%], PSNR in [dB], and $Q$ is 
	the JPEG compression level).
}
\label{tab:adaptive_jpeg_distance_appendix}

\end{minipage}

\end{table}

\hphantom{\\\\\\\\\\}

\end{document}